\documentclass[namedreferences]{SolarPhysics}
\usepackage[optionalrh]{spr-sola-addons} 
\usepackage{graphicx}        
\usepackage{color}           
\usepackage{url}             

\newcommand{\etal}{{\it et al.}}

\begin{document}

\begin{article}

\begin{opening}

\title{EUV SpectroPhotometer (ESP) in Extreme Ultraviolet Variability Experiment
(EVE): Algorithms and Calibrations\\ {\it Solar Physics}}

\author{L.~\surname{Didkovsky}$^{1}$\sep
        D.~\surname{Judge}$^{1}$\sep
        S.~\surname{Wieman}$^{1}$\sep
        T.~\surname{Woods}$^{2}$\sep
        A.~R.~\surname{Jones}$^{2}$
       }
\runningauthor{Didkovsky et al.} \runningtitle{ESP}

   \institute{$^{1}$ University of Southern California, Space Sciences Center
                     email: \url{leonid@usc.edu} email: \url{judge@usc.edu}  email: \url{wieman@usc.edu}\\
              $^{2}$ University of Colorado at Boulder, Laboratory for Atmospheric and Space Physics
                     email: \url{tom.woods@lasp.colorado.edu}   email: \url{andrew.jones@lasp.colorado.edu}
\\
             }

\begin{abstract}
The Extreme ultraviolet SpectroPhotometer (ESP) is one of five channels of the Extreme ultraviolet Variability Experiment (EVE) onboard the NASA {\it Solar Dynamics Observatory} (SDO). The ESP channel design is based on a highly stable diffraction transmission grating and is an
advanced version of the Solar Extreme ultraviolet Monitor (SEM), which has been successfully observing solar irradiance onboard the {\it Solar and
Heliospheric Observatory} (SOHO) since December 1995. ESP is designed to measure solar Extreme UltraViolet (EUV) irradiance in four first order bands of the diffraction grating centered around 19 nm, 25 nm, 30 nm, and 36 nm, and in a soft X-ray band from 0.1 to 7.0~nm in the zeroth order of the grating. Each band's detector system converts the photo-current into a count rate (frequency). The count rates are integrated over 0.25~sec increments and transmitted to the EVE Science and Operations Center for data processing. An algorithm for converting the measured count rates into solar irradiance and the ESP calibration parameters are described. The ESP pre-flight calibration was performed at the Synchrotron Ultraviolet Radiation Facility of the National Institute of Standards and Technology. Calibration parameters were used to calculate absolute solar irradiance from the Sounding Rocket flight measurements on 14 April 2008. These irradiances for the ESP bands closely match the irradiance determined for two other EUV channels flown simultaneously, EVE's Multiple Euv Grating Spectrograph (MEGS) and SOHO's Charge, Element and Isotope Analysis System / Solar EUV Monitor (CELIAS/SEM).

\end{abstract}
\end{opening}

\section{Introduction}
     \label{S-Introduction}
The {\it Solar Dynamics Observatory} (SDO) is the first NASA Living With a Star mission with its launch planned for February 2010. It will provide accurate
measurements of the solar atmosphere characteristics with high spatial and temporal resolution at many wavelengths simultaneously. These measurements will help
us understand the solar activity cycle, the dynamics of energy transport from magnetic fields to the solar atmosphere, and the influences of this
energy transport on the Earth's atmosphere and the heliosphere. Dynamic changes of the solar radiation in the extreme ultraviolet (EUV) and X-ray
regions of the solar spectrum are efficient drivers of disturbances in the Earth's space weather environment. The Extreme ultraviolet
Variability Experiment (EVE) is one of three instrument suites on SDO. EVE measures the solar EUV irradiance with unprecedented spectral
resolution, temporal cadence, accuracy, and precision. Furthermore, the EVE program will incorporate physics-based models of the solar EUV
irradiance to advance the understanding of solar dynamics based on short- and long-term activity of the solar magnetic features \cite{Woods06,
Woods09}.

ESP is one of five channels \cite{Woods06} in the EVE suite. It is an advanced version of the SOHO/CELIAS Solar Extreme ultraviolet Monitor (SEM)
\cite{Hovestadt95, Judge98}. SEM measures EUV solar irradiance in the zeroth diffraction order (0.1 to 50.0~nm bandpass) and two (plus and minus) first order diffraction
bands (26 to 34~nm bandpasses) centered at the strong He II 30.4~nm spectral line. More than 13 years of EUV measurements have shown that the SEM is a
highly accurate and stable EUV spectrometer \cite{Judge08} that has suffered only minor degradation, mainly related to deposition of carbon on
the SEM aluminum filters.

The ESP design is based on a highly stable diffraction transmitting grating \cite{Schattenburg90, Scime95}, very similar to the one used in SEM. ESP has filters, photodiodes, and
electronics with characteristics (transmission, sensitivity, shunt resistance, {\it etc.}) that are susceptible to some change over the course of
a long mission. Because of such possible degradation ESP shall be periodically calibrated throughout the mission. The calibration program includes a pre-flight
calibration followed by a number of sounding rocket under-flights with a nominally identical prototype of the flight instrument that is typically
calibrated shortly before and shortly after each under-flight. Calibration of the SDO/EVE and EVE/ESP soft X-ray and EUV flight instruments was performed at the National Institute of Standards
and Technology (NIST) at the Synchrotron Ultraviolet Radiation Facility (SURF-III). NIST SURF-III has a number of Beamlines (BL) with different
specifications and support equipment ({\it i.e.} monochromators, translation stages and calibration standards). The beamline used for a given
instrument depends on the instrument's characteristics and calibration requirements such as spectral ranges, entrance aperture,
alignment, {\it etc.} Before the flight ESP was integrated into the EVE package, it was first calibrated on SURF BL-9. BL-9 is equipped with a monochromator
capable of scanning through a wide EUV spectral band which allows the instrument efficiency profile for ESP to be measured in increments of 1.0~nm
over a spectral window of about 15 to 49~nm. After ESP was mounted onto EVE, the second ESP calibration was performed on SURF BL-2
\cite{Furst93}. BL-2 illuminates the instrument with the whole synchrotron irradiance spectrum, which can be shifted in its spectral range by changing the energy of the electrons circulating in the electron storage ring producing the synchrotron radiation. The intensity of the EUV beam is controlled by the current (and therefore number) of circulating
electrons. Because the EUV beam consists of a wide spectrum of photon energies, this type of calibration is a radiometric calibration. A major
part of this paper is related to the results of the ESP BL-9 and BL-2 calibrations.

 The scientific objectives of ESP are given in Section 2. Section 3 presents a short overview of the ESP channel. Section 4 describes an algorithm
 to calculate solar irradiance from count-rates measured on each of ESP's photometer bands, measurement conditions ({\it i.e.} ESP temperature,
 dark current, band response, {\it etc}.), ESP calibration data and orbit parameters. Section 5 shows results from ESP ground tests. Section 6 summarizes ESP pre-flight calibration results. A comparison of ESP solar irradiance measurements from the sounding rocket flight of 14 April 2008 with data from other EUV channels (EVE/MEGS sounding rocket instrument and SOHO/SEM) are given in Section 7. Concluding remarks are given in Section 8.

\section{ESP Scientific Objectives} 
      \label{S-ESP-Science}
      The scientific objectives of ESP were developed as a result of analysis of soft X-ray and EUV observations by diode based photometers with
      thin-film metal
      filters, including those on sounding rocket flights \cite{Ogawa90}, \cite{Judge98}, and TIMED/XPS \cite{Woods98a}. We found several critical limitations with those observations including significant spectral contamination in the filter based photometers, low duty cycle due to a low Earth orbit, and relatively low time cadence. Some of these limitations have been partially overcome with the SOHO/SEM instrument. More than 13 years of practically uninterrupted SEM observations have resulted in significant improvements in solar
      modeling with the use of the new $S_{10.7}$ solar index from SEM 26 to 34~nm bands \cite{Tobiska07}, in the Earth's atmosphere neutral density
      models \cite {Bowman06a, Bowman06b}, in studying ionosphere responses to the solar storms, see {\it e.g.}, \opencite{Tsurutani05}.
      However, some other limitations remain in the SEM observations, such as occasional energetic particle contamination of the EUV measurements, and a
      limitation on the data transfer rate, which results in a SEM time cadence of 15~s in the SOHO data
      stream compared to the instrument's native time resolution capability of 0.25 sec. Totally free of these limitations, ESP has the following scientific (and technical) objectives:\\
       - To provide 40 times better time cadence than SOHO/SEM in order to address solar dynamics at the highest EUV cadence to date. \\
       - To provide near real-time calibrated solar irradiance with minimal latency for rapid space weather prediction.\\
       - To provide much greater EUV spectral coverage than SOHO/SEM with 3 times as many data bands permitting atmospheric source region variability studies. ESP's four first order bands are centered at wavelengths of about 19, 25, 30,
       and 36~nm. These wavelengths contain strong solar EUV spectral lines traditionally used to study the dynamics of the solar atmosphere,
       {\it e.g.}, SOHO/EIT has 19.5~nm and 30.4 nm bands, TRACE has a 19.5~nm band, SDO/AIA has 19.3~nm, and 30.4~nm bands. These wavelengths, as well as the 0.1 to 7.0 nm bandpass
       observed with ESP's zeroth order quad diode band are in high demand for space weather applications, as proxies of the changes of the Earth atmosphere neutral
       density and ionization, and for building more accurate EUV solar models. In addition, the relative activity of the solar E, W, N, and S quadrants on the solar disk and their time dependence will provide data on large scale relative variability `quadrant to quadrant'.\\
       - To provide significantly more accurate EUV measurements in order to search for short and long term variability and its relationship to solar produced weather variability in the Earth's atmosphere. ESP has a more accurate method for compensating for changes in the measured signal related to changes of observing conditions than any other instrument of its class flown to date. This compensation includes a signal correction for changes in
      detector temperature, for changes of sensitivity due to radiation damage to the electronics, for changes in the amount of
      and residual sensitivity to scattered and stray visible light, and for contamination of the EUV signal due to energetic particles.

      All of these factors will significantly improve the accuracy of ESP observations. The scientific objectives of ESP complement those of EVE \cite{Woods09} with ESP providing long-term stable, photometrically accurate, high time cadence EUV measurements while the high resolution spectra in the broad wavelength range will be measured by the EVE/MEGS.
      ESP calibrated irradiance will be used as reference data for the EVE MEGS-A, MEGS-B, and SAM channels.

      In addition to the science objectives related to the EUV observations, ESP will detect fast changes of Solar Energetic Particle (SEP)
      fluxes associated with strong solar storm events. Geometrically accurate opto-mechanical models
      of the SEM and ESP were developed \cite{Didkovsky07a, Didkovsky07b} to determine the detector's sensitivity to the SEPs by modeling the
      stopping power for high-energy protons. The ESP stopping power model shows
      that a differential signal from the ESP detectors may be used to extract solar proton spectra in an energy range of 38 to 50~MeV with Full Width at Half Maximum (FWHM) of about 6~MeV.

      High-cadence and accurate EUV observations provided by the ESP will
      improve our knowledge of the energy release spectrum during solar flares, will increase the accuracy of EUV solar models including their
      prediction capabilities, add to the information about particle initiation and acceleration, and contribute to studies of the ionosphere and to Earth's
      atmosphere neutral density models.

\section{ESP Overview}
     \label{S-Overview}

SDO/EVE/ESP is an advanced version of the Solar Extreme ultraviolet Monitor (SEM) \cite{Judge98} that has been successfully working onboard
SOHO as part of the CELIAS experiment since 1995. A comparison of design features for SEM and ESP is shown in Table 1.
\begin{table}[h]
\caption{A comparison of SEM and ESP design features }
\begin{tabular}{|c|c|c|} 
\hline
  Design Feature & SEM & ESP \\
\hline
Number of bands  &  3 & 9 \\
Filter wheel  &  No &  Yes, with 5 filters \\
Dark band  &  No & Yes \\
Quad diode bands  &  No & Yes \\
Dark mode for all bands  &  No & Yes \\
Gain reference mode  &  No & Yes \\
\hline
\end{tabular}
\end{table}
Each of the ESP features shown in Table 1 that were not incorporated into SEM are enhancements which ultimately improve the accuracy and performance of the channel and thus, the quality of the data it contributes to the science of solar EUV variability. For example, the gain reference mode is an important feature to track changes in each band's gain during the flight lifetime. These changes may be related to both ESP electronics temperature and electronics degradation, {\it e.g.}, due to the Total Ionizing Dose (TID) radiation damage even though great care was taken to mitigate such change.

The ESP instrument configuration is shown in Figure 1. A high-voltage grid in front of ESP, with a potential of 1000~V, eliminates low-energy charged particles that could enter ESP through the entrance slit (1 $\times$ 10~mm) and deposit energy in the ESP detectors creating additional output signal beyond that due to EUV photons.
The baffles are designed to minimize the amount of visible (scattered and stray) light that reaches the detectors. The limiting apertures on the detectors define the edges of the spectral bands.
  \begin{figure}[ht]
   \begin{center}
   \begin{tabular}{c}
   \includegraphics[height=8.5 cm]{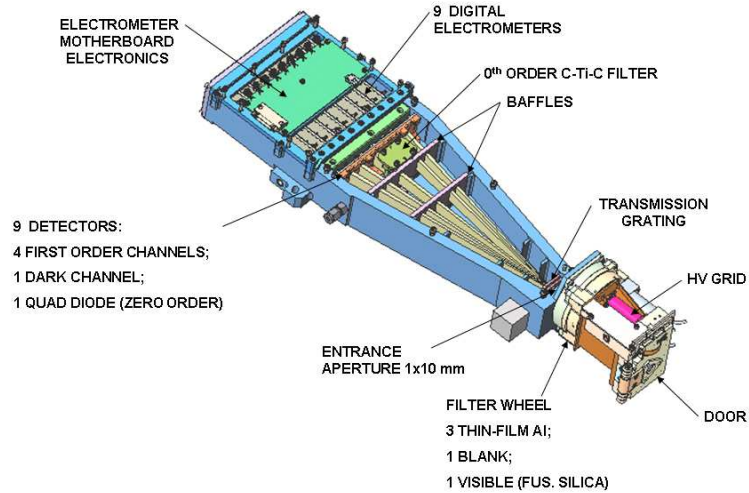}
   \end{tabular}
   \end{center}
   \caption[Figure 1]
   { \label{fig:Figure 1}
An overview of the ESP configuration (see details in the text).   }
   \end{figure}
A filter wheel mechanism in front of the entrance aperture has three aluminum filters (one primary and two spare), a visible light (fused silica) filter, and a dark filter. The thin-film aluminum filter rejects the visible light and transmits the EUV radiation to the bands.
The fused silica filter is used from time to time to determine the amount of visible light that may reach the EUV detectors and should be
subtracted from the EUV band signals. The dark filter is to measure dark currents on each band directly.

The detector assembly has nine detectors including four first order bands, four zeroth order bands in the form of a Quadrant Diode (QD), and a dark band with a diode that is constantly closed off. Nine digital electrometers
convert diode currents into count-rates using Voltage-to-Frequency Converters (VFC). The electrometer motherboard electronics provides
synchronization of operations and is the connection point of ESP to the EVE electronics.

The first order bands are centered around 19~nm, 25~nm, 30~nm, and 36~nm. They each have a spectral resolution of about 4~nm at FWHM.

The quad pattern of the zeroth order bands is centered about the optical axis of ESP. These bands provide information about de-centering ESP from misalignment or
pointing errors (see Section 6.4.1). A non-uniformity in the solar disk irradiance distribution related to the dynamics of Active Regions, ({\it e.g.} the growth
or decay of such regions or their position changes due to solar rotation) will be detected by the QDs. The signal non-uniformity across the QD bands will be used to determine an angular correction on the first order band signals, and be used as a real-time solar flare detector. For normal in-flight operation solar irradiance to the zero-order band goes through an input aluminum filter installed in the filter
wheel and through a free-standing C-Ti-C filter installed between the diffraction grating and the QD detector. These filters constrain the spectral bandpass of the zero-order band, which is from about 0.1 to 7.0~nm at greater than 10\% efficiency.

The dispersive element in ESP is a free-standing gold foil transmission grating with a 200~nm period. Gratings of this type have been used on multiple satellite and sounding rocket flights and have proven to be reliable for solar EUV irradiance measurements even on long-term space missions. The long-term accuracy of ESP measurements depend on information about possible changes
of ESP characteristics such as filter transmissions, detector responsivity, and electronics sensitivity. This information will be provided
by regular under-flight calibrations using an ESP sounding rocket clone instrument. In addition to these under-flight calibrations a number of methods are in place for monitoring and maintaining ESP calibration onboard. These methods include: \\
 - In-flight compensation for thermal changes of dark currents, see Equations (3 to 8) in Section 4;\\
 - Determination and subtraction of signal related to the scattered visible light, see Equations (9, 10) in Section 4;\\
 - Determination and correction for guiding errors (tilts), see Subsection 6.4.1;\\
 - Correction of the measured whole disk solar EUV irradiance for the location of solar flare, see Subsection 6.4.1;\\
 - Determination and subtraction of energetic particle related signals, see Equation (2) in Section 4;\\
 - Use of spare input aluminum filters in the event of filter damage ({\it e.g.} pin-hole damage), see Subsection 6.4.3 and Figure 16;\\
 - Minimizing TID radiation damage to electronics through the use of an appropriately thick ESP housing and strategically located spot shielding;\\
 - A reference mode for each ESP band to determine and correct possible changes of electronics due to the TID, see Equation (1) in Section 4 and Figure 5.\\

 All these methods along with sounding rocket under-flights will guarantee highly reliable and accurate EUV measurements by ESP.

\section{An algorithm to convert ESP count rates into solar irradiance}
     \label{S-Algorithm}

EUV solar irradiance is detected by the ESP band detectors as an increase in diode current beyond a base level dark current that is intrinsic to the diode / electrometer combination and measured even in the absence of photon illumination. These combined (dark + signal)
currents are converted by the electrometers into voltages and then into discrete pulses with frequencies that are linear functions of the input signals. The frequencies are measured as counts per a fixed time interval, {\it e.g.} per
0.25~sec. They are also referred to as count rates. The count rate of each band is a function of many input parameters, {\it e.g.}, EUV
solar intensity and spectral distribution of solar irradiance, dark currents, sensitivity to different wavelengths, distance to the Sun,
degradation of ESP, {\it etc.} An algorithm to convert ESP count rates into solar irradiance has been developed. It combines all these
input parameters into equations described below. Some of the ESP input parameters were determined based on ground tests. Parameters in this group, such as dark currents, reference voltage levels, and thermal changes of
these parameters, do not require any input radiation. The parameters of the second group are determined based on the ESP calibration at the
NIST SURF-III. These parameters include ESP
sensitivity to the angular position of the input source of irradiation, the spectral efficiency of each band, determination of the spectral
bandpass profiles, influences from the higher orders of the grating, characteristics of the thin-film filters, {\it etc.} The parameters of the third group
are related to solar observations and will be determined or corrected during the mission time. These parameters include a contribution of the visible light and energetic particles to the ESP band
signals, a correction of the EUV signal due to changes in the dark currents, and a correction for ESP degradation based on the comparison of the
orbit measurements with the sounding rocket under-flight measurements made simultaneously.

Solar irradiance $E_{i}$ in units of $W/m^{2}$ for each scientific band $i$ is determined as
\begin{equation}    
E_{i}(\lambda,t) = \frac {C_{i, eff}[1-\frac {dG_{i}(T,V,TID)}{\Delta{t}}]} {{A}\frac{\int_{\lambda0 -
\Delta\lambda}^{\lambda0+\Delta\lambda}R_{i}(\lambda,\alpha,\beta)\frac{\lambda}{hc} F_{i}(\lambda)d\lambda}{\int_{\lambda0 -
\Delta\lambda}^{\lambda0+\Delta\lambda}F_{i}(\lambda)d\lambda}f_{i,degrad}(t)f_{1AU}(t)} -E_{OS}
\end{equation}
where $C_{i, eff}$ in the ESP Equation (1) is the band's effective count rate. For the time $t_{j}$
\begin{equation}    
C_{i, eff}(t_{j}) = C_{i, meas}(t_{j}) - C_{i, ch.dark}(t_{j}) - C_{i, particleBG}(t_{j}) - \Delta C_{i, vis}(t_{j}),
\end{equation}
which, for each ESP band $i$, is the result of subtracting the background signal (signal not related to EUV within the designed
band-passes) from the total measured signal $C_{i, meas}$. The background signal is composed of three parts: dark count rate $C_{i, ch.dark}$,
measured when the dark filter is in position, an energetic particle related signal $C_{i, particleBG}$, and a difference $\Delta C_{i, vis}$
between the visible scattered light count rate when the visible (fused silica) filter is in place and the dark count rate. We obtain all three of these contributors to the background signal based on ESP design parameters, results of ground tests, and laboratory measurements as described below. All these contributors will be measured in flight.

One of the nine ESP bands, a dark band $C_{dark}$, is completely closed at all times to prevent light from reaching it. It is possible
to use the $C_{dark}$ measurements (available at any time) as a proxy for dark counts in the science bands, and as a proxy for particle
background signal. The use of the dark band counts $C_{dark}$ saves time during mission operations by reducing the frequency with which one
needs to move the filter wheel back and forth between the dark filter and the primary observing (aluminum) filter. ESP ground tests showed that
$C_{i, ch.dark}$ is sufficiently stable over time periods for which the temperature $T$ is stable. We may assume, therefore, that  counts,
$C_{i, ch.dark}(t_{j})$, at a given time, ($t_{j}$), are approximately equal to those, $C_{i, ch.dark}(t_{1})$, measured at the previous time,
($t_{1}$), that the dark filter was in place  and that the difference between the two is related to the temperature change $\Delta T_{j}$
determined from the ground tests
\begin{equation}    
C_{i, ch.dark}(t_{j},T_{j}) = C_{i, ch.dark}(t_{1}, T_{1}) + \Delta C_{i, ch.dark}(t_{j}, \Delta T_{j}),
\end{equation}
Where:
\begin{equation}    
\Delta C_{i, ch.dark}(t_{j},\Delta T_{j}) = C_{i, ch.dark}(t_{j}, T_{j}) - C_{i, ch.dark}(t_{1}, T_{1})
\end{equation}
Then, the change in a band's dark count rate may be replaced with the dark band measurements $C_{dark}$ and the band's dark proxy
counts $C_{i, chd.proxy}$
\begin{equation}    
\Delta C_{i, ch.dark}(t_{j},\Delta T_{j}) = \frac {C_{dark}(t_{j}, T_{j})} {C_{i, chd.proxy}(T_{j})}-C_{i, ch.dark}(t_{1}, T_{1}) ,
\end{equation}
Thus, a band's dark count rate for the time $t_{j}$ is determined using previous, {\it e.g.}, a couple of days earlier, measurements with the dark
filter in place for the time $t_{1}$ and the band dark proxy, where:
\begin{equation}    
C_{i, chd.proxy}(T_{j}) = \frac {C_{dark}(T_{j})} {C_{i, ch.dark}(T_{j})},
\end{equation}
where $ C_{dark}(T_{j})$ and $C_{i, ch.dark}(T_{j})$ are determined for a working range of temperatures $T$ and tabulated during ground tests.
With all the above considerations, the relation for the $C_{i, ch.dark}(t_{j})$ in Equation 2 is
\begin{equation}    
C_{i, ch.dark}(t_{j}) = \frac {C_{dark}(t_{j},T_{j})} {C_{i, chd.proxy}(T_{j})}
\end{equation}
Another method to determine the $C_{i, ch.dark}(t_{j})$ is to use a change of the dark count
rate as a function of the detector temperature $T_{j}$ determined from the thermal ground tests for each band $i$
\begin{equation}    
C_{i, ch.dark}(t_{j})=F_{i,ch.dark}(T_{j})
\end{equation}
One of these two methods, either the dark band proxy (Equation 7) or the function of change of the band's dark count rate (Equation 8) will be chosen as the
most accurate method after engineering measurements on orbit and will be used in Equation 2.

The degree to which each detector is shielded from energetic particles varies slightly depending on its position within the housing
({\it i.e.} its location with respect to housing walls and other mechanical structures). Because of this positional dependence of shielding thickness
and, accordingly, total stopping power for energetic particles, the band responses to energetic particle flux were studied
\cite{Didkovsky07b} for both isotropic and concentrated fluxes. This analysis, based on a geometrically accurate 3-dimensional (SolidWorks)
model showed that the energy deposited in the ESP detectors $C_{i, particleBG}$ by particles associated with a quiet Sun or with small solar
storm events is so small that it is less than the noise. For solar storms
associated with extreme solar flare events the particle related band signal may be determined with corresponding position dependent modeled
corrections to the signal extracted from the dark band.

The term $\Delta C_{i, vis}$ in Equation 2 shows the portion of the band's signal related to scattered visible light. Laboratory
measurements, in which a strong visible light source was placed in front of the ESP entrance aperture (with the  aluminum filter removed),
showed very little to no visible light sensitivity in any of the ESP first order bands. Nevertheless, in case of some changes the conditions of
visible light scattering, the updated $\Delta C_{i, vis}$ may be found as
\begin{equation}    
\Delta C_{i, vis}(t_{j}) = \frac{C_{fus.silica}(t_{j}) - C_{i, ch.dark}(t_{j}) - C_{i, particleBG}(t_{j})}{T_{fus.silica} +\Delta
T_{fus.silica}} (\frac{1 AU}{f_{1 AU}})^2
\end{equation}
where $T_{fus.silica}$ and $\Delta T_{fus.silica}$ are, respectively, the transmission of the fused silica filter and its change from pre-flight
to flight conditions (Equation 10). The amount of incoming visible light reaching the ESP entrance aperture is corrected by the relative distance to the Sun
squared.
\begin{equation}    
\Delta T_{fus.silica}= T_{fus.silica flight} - T_{fus.silica preflight}
\end{equation}
When the visible light signal, measured with the fused silica filter, is smaller than the sum of the band's dark and particle background
signals, then $\Delta C_{i, vis}$ is set to zero.

Temporal changes of the band's gain $dG(T,V,TID)/\Delta{t}$ are caused by the changes of the following variables, temperature $T_{i}$, the
reference voltage $V_{i}$ used to calibrate the electrometer's VFC, and the TID, which
can affect the parameters of the VFC. If the reference voltage $V_{i}$ is quite stable against TID and its temperature changes are determined
during ground tests, in-flight changes of the gain after subtraction of the temperature/voltage changes give us information about the TID
influence.

In Equation 1, ($A$) is the ESP entrance aperture area, $R(\lambda,\alpha,\beta)$ is the band
responsivity to the wavelengths within the range of the band's spectral window, which
is also a function of angular alignment, ($\alpha$ and $\beta$), and the solar field of view
(FOV). During solar observations, each channel's bandpass is expanded compared
to the calibration bandpass. The band responsivity $R(\lambda,\alpha,\beta)$ in Equation 1 is
the result of convolving the responsivity profile measured at BL-9 $\xi_{i}(\lambda)$
and the band's exit slit function $S_{\beta}(\lambda, FOV)$:
 \begin{equation}    
R(\lambda,\alpha,\beta)= S_{\beta}(\lambda, FOV) * \xi_{i}(\lambda),
\end{equation}
Figure 2 shows an example of the convolution for Ch8.
  \begin{figure}[ht]
   \begin{center}
   \begin{tabular}{c}
   \includegraphics[height=5.5 cm]{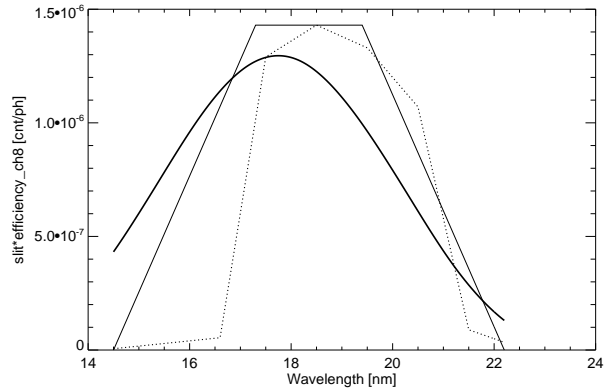}           
   \end{tabular}
   \end{center}
   \caption[Figure 2]
   { \label{fig:Figure 2}
   An example of the convolution (thick line) for the Ch8 efficiency profile (dotted line) and
exit slit trapezoidal function (thin line).}
   \end{figure}
The relative amplitude of the spectral
radiation component $F(\lambda)$ shown in Equation 1 is $F(\lambda)d\lambda$ divided by $\int_{\lambda0 -
\Delta\lambda}^{\lambda0+\Delta\lambda} F_{i}(\lambda) d\lambda$. $f_{degradation}(band,t)$ is the time and band dependent
degradation factor and $f_{1AU}$ is a measure of the distance from ESP to the Sun in A.U.

The term $E_{OS}$ is the portion of the band's solar irradiance measurement $E_{i}(\lambda,t)$ that is due to shorter wavelengths at higher
diffracted orders.
\begin{equation}    
E_{OS}= \sum_{n=2,3} R_{OS}E_{n},
\end{equation}
where $R_{OS}$ is the band's relative sensitivity to higher orders 2 and 3, $R_{OS} = R_{n}/R_{1}$. $E_{n}$ is the portion of energy that
reaches the band's detector from a higher order.

\section{Results from the Ground Tests}
     \label{S-Ground-tests}
 Ground tests performed on ESP (as part of the EVE assembly) included thermal-vacuum (TV) tests, electro-magnetic interference (EMI), and electro-magnetic compatibility (EMC) tests. Concurrent EMI and EMC tests
 were performed and produced some increased (peak) count rates compared to the smooth changes of currents with temperature observed without EMI/EMC. Analysis of thermal changes of both dark count rates and reference count rates described in this Section was first applied to the whole set of data, which includes the EMI/EMC peaks. A refined data set which excludes peaks related to `survival condition' electromagnetic fields far exceeding those expected under operational conditions, led to similar (within the standard errors) characteristics of the thermal current changes. This refined set of measurements was used to build the thermal current curves for each ESP band.
     \subsection{Thermal changes of dark count rates}
     \label{S-Dark-tests}
 A variety of dark current changes (dark count rates) for each of nine ESP bands is shown in Figure 3. All the thermal characteristics were approximated with a third degree polynomial fit to about $5.2\times10^{6}$ measurements obtained during TV tests in 2007. The standard deviation error for these fits are within $\pm$ 0.5 cnt/0.25~s with the mean number of $\pm$ 0.38 cnt/0.25~s.

  \begin{figure}[ht]
   \begin{center}
   \begin{tabular}{c}
   \includegraphics[height=5.5 cm]{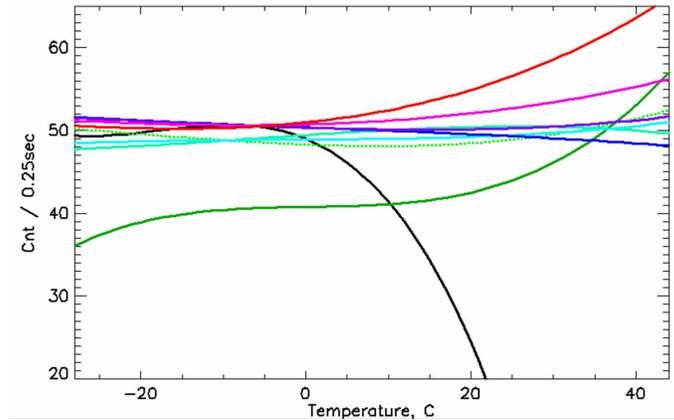}           
   \end{tabular}
   \end{center}
   \caption[Figure 3]
   { \label{fig:Figure 3}
Thermal changes of ESP dark count rates in the temperature range of -$28^{\circ}$ C to $34^{\circ}$~C. The bands are marked with different colors, Ch1 (black), Ch2 (dark green), Ch3 (dotted), Ch4 (light green), Ch5 (cyan), Ch6 (dark blue), Ch7 (violet), Ch8 (pink), and Ch9 (red).   }
   \end{figure}
 Figure 3 shows that all ESP band dark thermal changes for the temperature range from -$28^{\circ}$~C to $34^{\circ}$~C, except for Ch1, are within a range of count rates of about $\pm$ 15 cnt/0.25~s. Ch1 thermal changes are much larger for the positive temperatures. However, with the ESP detector planned temperature of about $10^{\circ}$~C on orbit, dark count rate for Ch1 will be suitable for subtraction as long as it remains positive. One-sigma standard errors for the Ch1 thermal changes are shown in Figure 4.
  \begin{figure}[ht]
   \begin{center}
   \begin{tabular}{c}
   \includegraphics[height=5.0 cm]{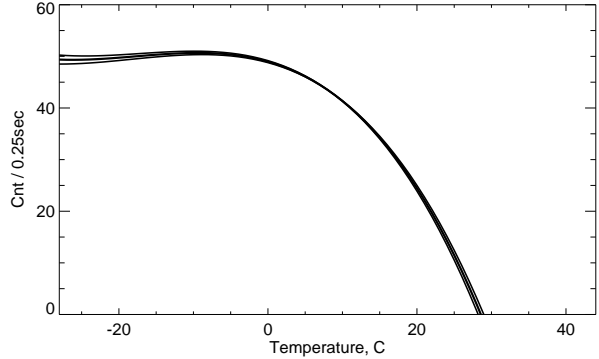}
   \end{tabular}
   \end{center}
   \caption[Figure 4]
   { \label{fig:Figure 4}
One-sigma standard errors for ESP Ch1 thermal changes. Each point measurement errors were assumed as $\pm 0.5$ cnt/0.25~s.}
   \end{figure}
The thermal changes of dark counts determined during 2007 will be verified and corrected if required during engineering tests on orbit.

     \subsection{Thermal changes of reference count rates}
     \label{S-Reference-tests}

The reference mode is designed to determine possible changes of the electronics gain of each VFC. When the reference mode
is enabled by software, a stable reference voltage is applied to the outputs of the electrometers and, thus, to the inputs of the VFCs. Even with the thick ESP envelope and additional metal shields at the VFC component locations, the VFC may still be susceptible to some degradation due to the X-ray and particle radiation environment, which is measured as accumulated TID. Periodic measurements of the count rates in reference mode and comparison of these count rates to the pre-flight values will show these TID related changes. In addition to possible changes of count rates related to TID, reference mode count rates are sensitive to the changes of electronics temperature. Figure 5 shows these thermal changes determined from the 2007 TV tests.
  \begin{figure}[ht]
   \begin{center}
   \begin{tabular}{c}
   \includegraphics[height=5.0 cm]{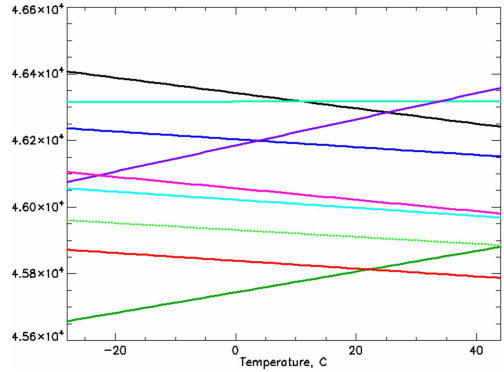}
   \end{tabular}
   \end{center}
   \caption[Figure 5]
   { \label{fig:Figure 5}
Thermal changes of ESP reference count rates in the temperature range of -$28^{\circ}$~C to $34^{\circ}$~C for ESP bands from Ch1 to Ch9. The color spectrum is the same as in Figure 3. The vertical axis shows counts per 0.25 s.    }
   \end{figure}
The thermal changes of the counts in the reference mode show small linear trends and these changes will be applied to the correction of the electronics gain (see Equation 1, term $dG_{i}(T,V,TID)$.

\section{ESP pre-flight calibration}
     \label{S-Pre-flight}
 The goal of the pre-flight calibration is to determine all the channel variables that affect the calculation of solar irradiance (Equation 1).
 Some of these variables may be directly measured during ground tests, others are determined from the SURF calibrations. A spectrum of energetic particle energies that are sufficient to create a particle background signal on the ESP detectors is determined from the ESP opto-mechanical proton interaction model.

    \subsection{ESP variables: where they are measured and how accurately}

 Equation (1) for solar irradiance shows all ESP variables (parameters) that need to be determined for converting measured count rates into
 corresponding irradiance. Each of these variables is shown in Table 2 with an indication of the Equations 1 through 11 in which it appears (first column). The second column shows the method of determination, {\it e.g.}, $G$ is for ground tests (non-SURF), and the third column shows a relative statistical error of determination.
\begin{table}[h]
\caption{ESP variables: where they are determined (second column) and how accurately (third column) }
\begin{tabular}{|c|c|c|} 
\hline
  ESP parameter and (Equation) & Where it is determined & Relative error, \% \\
\hline
$C_{i, ch.dark}$ (2-9)  &  G & 0.5 \\
$C_{dark}$ (5-7)& G, SURF &  0.5 \\
$C_{i, chd.proxy}$ (5-7)  &  G & 0.7 \\
$C_{i, particleBG}$ (9)   & Model & 0.5 -- 5.0$^{1}$ \\
$\Delta C_{i, vis}$ (2,9)  & G, SURF & 0.0 -- 5.12$^{2}$\\
$C_{fus.silica}$ (9) & G, F & 0.0 -- 0.5$^{3}$ \\
$T_{fus.silica}$ (9,10) & G, F & 0.5 \\
$\Delta T_{fus.silica}$ (9,10)& G, F & 0.5 \\
$C_{i, meas}$ (2) & G, SURF, F & 0.5 \\
$C_{i, eff}$ (2) & F & 0.87 -- 7.2$^{4}$ \\
$dG_{i}(T,V,TID)$ (1) & G, F & 0.7 \\
$R_{OS}$ (11) & SURF & 4.2 \\
$E_{OS}$ (1,11) & SURF & 5.9 \\
$ A $ (1) & G & 0.1 \\
$R_{i}(\lambda,\alpha,\beta)$ (1) & SURF & 2.5 \\
$\int R_{i}(\lambda,\alpha,\beta) F_{i}(\lambda)d\lambda$ (1) & SURF & 9.8 \\
$\int F_{i}(\lambda)d\lambda $ (1) & SURF & 5.6 \\
$f_{i,degrad}$ (1) & F, UnderFlight, SURF & 15.9 \\
$f_{1AU}$ (1) & F & 0.1 \\
$E_{i}(\lambda,t)$ (1) & F & 21.3$^{4}$ \\
\hline
\end{tabular}
\end{table}
The notes in Table 2 are:\\
Note$^{1}$: The larger value is for extreme solar flare events \\
Note$^{2}$: The larger value is for the zeroth order bands \\
Note$^{3}$: The larger number is based on the assumption that there is some detectable amount of visible light\\
Note$^{4}$: The larger number is for a combination of extreme observing conditions\\

The largest source of uncertainty (error) is related to long term degradation trends (see $f_{i,degrad}$ in Table 2).
This error will be significantly reduced (to about 5\%) after 3 to 5 sounding rocket under-flights sufficient to understand and model the degradation curve. This reduced degradation error will decrease the uncertainty of determination of solar irradiance from 21.3\% shown in Table 2 to about 10\%.

     \subsection{ESP Calibration Overview}
     \label{S-calibration-overview}
 ESP calibrations at NIST SURF include initial calibration at the BL-9 where ESP was calibrated before assembling it to the SDO/EVE, and a radiometric calibration at BL-2 after assembling ESP to the EVE suite of channels. BL-9 is used primarily for detailed spectral profiles for each first-order band and filter transmission calibration and the SURF BL-2 is used primarily for end-to-end radiometric calibration of ESP.

 The BL-9 calibration at SURF consists of measurements of the channel response to a set of narrow wavelengths provided by a monochromator. BL-9 has a small vacuum tank with a translation stage where the instrument, {\it e.g.} ESP, is installed for testing. The tank provides ESP alignment through linear motions (X and Y) and tilts (Pitch and Yaw). Additionally, to reduce the influence of the beam linear polarization, ESP can be rotated (roll axis) to horizontal, vertical, or 45$^{\circ}$ orientations. The optical layout for the ESP BL-9 calibration is shown in Figure 6.
  \begin{figure}[ht]
   \begin{center}
   \begin{tabular}{c}
   \includegraphics[height=7.5 cm]{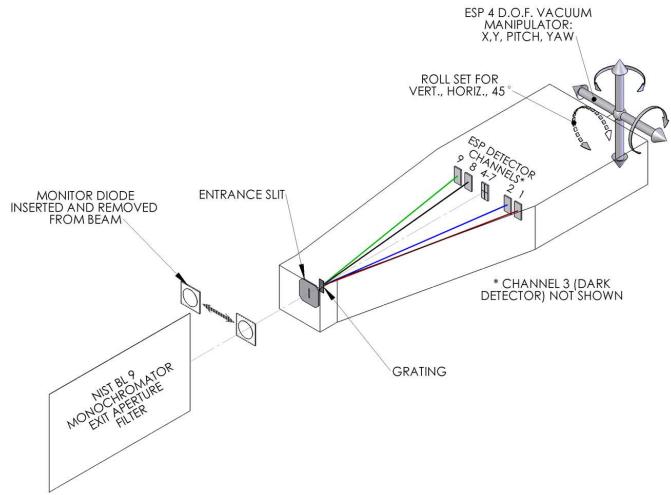}
   \end{tabular}
   \end{center}
   \caption[Figure 6]
   { \label{fig:Figure 6}
Opto-mechanical layout for ESP BL-9 calibration. The monochromator and BL-9 filters provide 27 test wavelengths over the range from 15.6 to 49.0~nm. A removable monitor diode measures the beam intensity before and after each wavelength's test run. ESP's orientation shown is horizontal. Note, the Al and Ti filters were removed from ESP during that BL-9 calibration. }
   \end{figure}
  A wavelength scan allowed determination of the spectral sensitivity profiles and efficiencies for each of the ESP first order bands. Due to the small reflectivity of the BL-9 grazing incidence diffraction grating at short wavelengths, the QD (zeroth order) bands were not calibrated at BL-9. The BL-9 vacuum tank mechanical dimensions are too small to permit insertion of the the whole flight ESP, {\it e.g.}, with the filter wheel, for calibration. To fit the allowed dimensions, the entrance door and the filter wheel with filters were removed from the flight ESP during BL-9 calibration.

 For the BL-2 radiometric calibration \cite{Didkovsky07c} the EVE suite of channels was installed in a large vacuum tank used for calibration of large NASA instruments. The intensity of BL-2 synchrotron light is accurately known to about 1\% and is a primary radiometric standard at NIST.  The key parameters for the synchrotron source are its beam energy and beam current, and these are provided by SURF who calibrates these parameters on a regular basis. SURF provides beam energy and current, X, Y, Pitch, and Yaw control. These input parameters together with the band's signal are recorded during calibration. The measurements are then used to determine the channel parameters (Table 2, marked as SURF in the second column). The optical layout for ESP BL-2 calibration is shown in Figure 7.
  \begin{figure}[ht]
   \begin{center}
   \begin{tabular}{c}
   \includegraphics[height=7.3 cm]{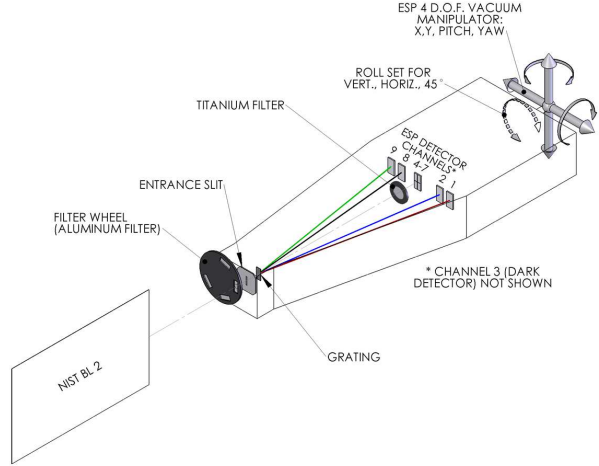}
   \end{tabular}
   \end{center}
   \caption[Figure 7]
   { \label{fig:Figure 7}
Opto-mechanical layout for ESP BL-2 calibration. ESP is assembled to the EVE but calibrated separately from other EVE channels when the ESP entrance slit and optical axis is aligned to the BL-2 optical beam. ESP's orientation shown is horizontal. }
   \end{figure}
The goal of ESP alignment to the beam is to make the ESP optical axis coincident with the beam optical axis. QD bands are usually used for ESP alignment. The alignment consists of a number of scans in two orthogonal directions (X and Y) to find the center of the beam. After finding the center, the Yaw and Pitch tilts are adjusted to have approximately zero X and Y coordinates, calculated from four QDs.

Because synchrotron light is highly polarized, the SURF calibrations usually need to be performed with two orthogonal orientations to account for the band sensitivity to polarization.  The ESP grating is sensitive to polarization but photometers / detectors are not.  ESP efficiency is higher for the orientation in which the grating grooves are aligned with the direction of the synchrotron beam polarization.  The gimbal table inside the BL-2 has a mounting ring for the instrument, and this ring can be rotated so the instrument has any angle relative to the synchrotron beam (polarization).  Any two orthogonal orientations can be used, such as -45$^{\circ}$ and +45$^{\circ}$.  However, it has been shown for several grating spectrometers at SURF that a single calibration can be done at 45$^{\circ}$, and the same result as averaging
horizontal and vertical calibrations is obtained.  The ESP pre-flight calibration is done at an orientation of 45$^{\circ}$.

     \subsection{Results from the BL-9 Calibration}
     \label{S-BL-9-calibration}
ESP was calibrated at BL-9 in two calibration modes, both horizontal ($H$) and vertical ($V$) orientations to the SURF beam. These provide a full set of information about sensitivity of ESP to the BL-9 linear polarization. Each of these calibration modes required different mounting of ESP to the tank's vacuum flange and was followed by the alignment of ESP to the beam optical axis. Determination of ESP band sensitivities to the angular position of a source of irradiation in the ESP FOV was provided by corresponding tilts of the ESP's aligned optical axis in the FOV range. The mean $(H + V)/2$ result of calibration was the band's efficiency $\xi$ determined as
\begin{equation}    
\xi (\lambda)_{0.25 sec} = \frac{(C_{ch} - C_{dark})}{F_{\lambda}}
\end{equation}
where $F_{\lambda}$ is
\begin{equation}    
F_{\lambda} = \frac{I_{monitor}}{1.602 \times 10^{-19} \times D_{d.eff}(\lambda)}
\end{equation}
where $I_{monitor}$ is the monitor diode current measured for the whole beam, $D_{d.eff}$ is the SURF calibrated (against an absolute radiometric standard) wavelength dependent quantum efficiency of the monitor diode. The size of the beam used for any of $H$ or $V$ orientations was decreased to the size smaller than the size of the ESP entrance aperture ($1 \times 10$~mm).

    \subsubsection{ESP filter transmissions}

    Efficiencies of the first order bands for the flight ESP were determined in 2006 from the BL-9 calibration with the aluminum filter removed (Figure 6). The correction of these efficiencies for corresponding wavelength dependent aluminum filter transmission were performed later, in 2008 after measurements at BL-9. ESP aluminum filters have a thickness of 150 $\pm$ 5~nm. The result of this measurement is displayed in Figure 8.
  \begin{figure}[ht]
   \begin{center}
   \begin{tabular}{c}
   \includegraphics[height=4.5 cm]{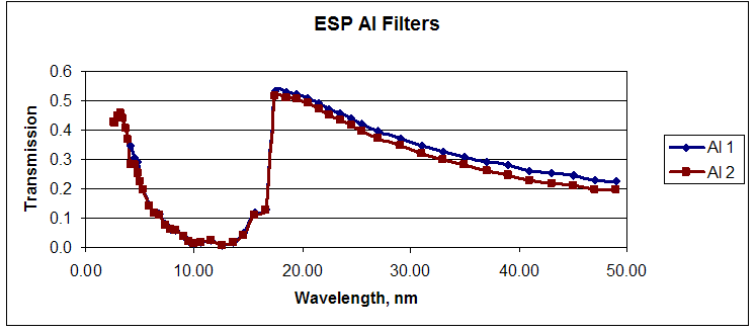}          
   \end{tabular}
   \end{center}
   \caption[Figure 8]
   { \label{fig:Figure 8}
ESP aluminum filters transmission as measured at BL-9. Both filters (Al 1 and Al 2 are from the same batch that was used for the flight ESP filters.}
   \end{figure}
    It shows a typical (small) deviation in the filter transmission related to two sources of uncertainty, the aluminum film thickness and the thickness of the oxide layers of the filters.

   The ESP C-Ti-C filter is a composite filter consisting of a 284~nm thick titanium layer between two carbon layers, each 19~nm thick. Figure 9 shows a comparison of the measured transmission of the zeroth order C-Ti-C filter (with the minimum wavelength limited by the BL-9 design to 3.65~nm) and transmission of a modeled C-Ti-C filter with 284~nm of Ti and 38~nm of C but without a support mesh structure. This model is based on calculations using the atomic parameters by \inlinecite{Henke93}.
  \begin{figure}[ht]
   \begin{center}
   \begin{tabular}{c}
   \includegraphics[height=5.0 cm]{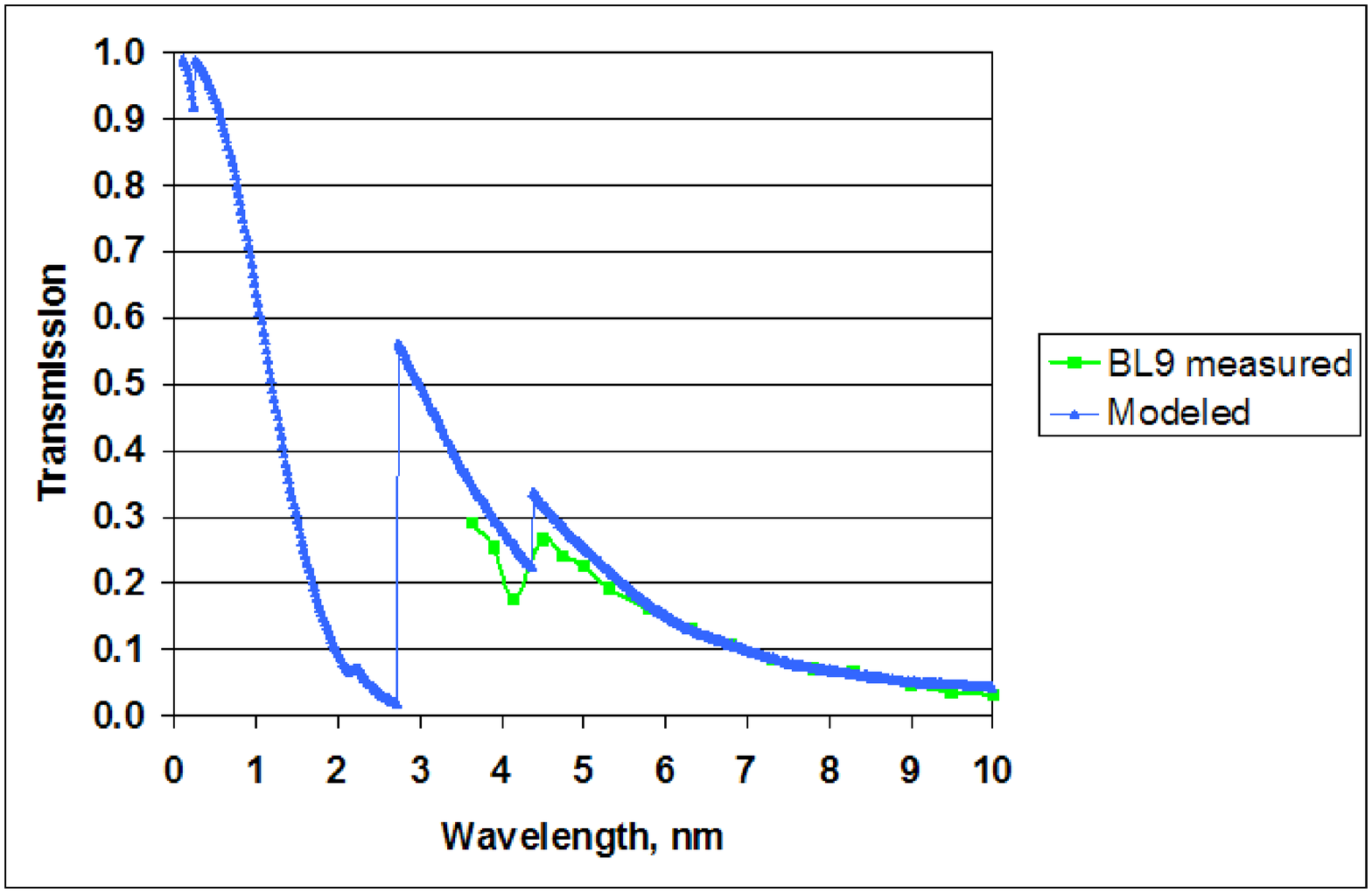}              
   \end{tabular}
   \end{center}
   \caption[Figure 9]
   { \label{fig:Figure 9}
A typical modeled transmission profile (blue line) of a C-Ti-C filter with a thickness of 284~nm in a wavelength range of 0.1 to 10~nm. This modeled transmission profile is compared to that of the ESP titanium filter as measured at BL-9 (green line) for a wavelength range of 3.65 to 49.0~nm. Figure 9 shows transmission for shorter wavelengths, from 10~nm. The ESP Ti filter is from the batch used for the flight ESP filter.}
   \end{figure}

     \subsubsection{Measured transmission of the diffraction grating}

The angle $\varphi_{m}$ between the perpendicular to the grating and the direction to the transmission maxima is a function of the angle between the ESP optical axis and the direction of the input beam. According to the theory of a diffraction grating, this angle $\varphi_{m}$ for normal incidence is a function of the wavelength $\lambda$ and the period of the grating $d$ (for the flight ESP $d$ = 201~nm):
\begin{equation}    
 d \times sin(\varphi_{m}) = m \times \lambda
\end{equation}
where $m$ is the diffraction order, $m$ = 0, ±1, ... If the input beam does not coincide with the ESP optical axis, {\it e.g.} is tilted, the output (diffracted) beam is shifted in direction:
\begin{equation}    
 d \times [sin(\theta)-sin(\varphi_{m})] = m \times \lambda
\end{equation}
where $\theta$ is the angle of the input beam with respect to the grating normal; $\varphi_{m}$ is direction to the maxima. Efficiency profiles for the tilted ESP positions will be analyzed in the corresponding section (6.4.1). Figure 10 shows transmission of the diffraction grating at 30.4~nm in the zeroth and first orders for the zero incidence angle beam measured at the SURF reflectometer.
  \begin{figure}[ht]
   \begin{center}
   \begin{tabular}{c}
   \includegraphics[height=6.2 cm]{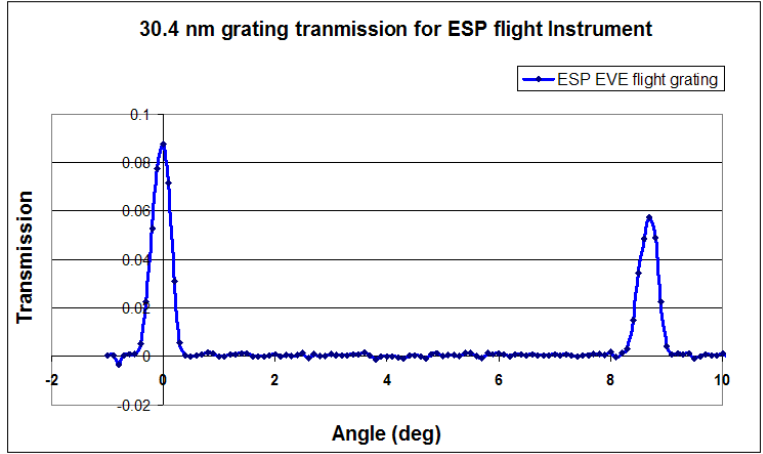}         
   \end{tabular}
   \end{center}
   \caption[Figure 10]
   { \label{fig:Figure 10}
SURF measurements of the transmission of the diffraction grating for Ch9 (30.4~nm) in the zeroth order (central peak) and the plus first order. }
   \end{figure}
The amplitude of the zeroth order transmission peak is about 9\%. The first order transmission peak at $\varphi_{m}$ = 8.7$^{\circ}$ (Equation 15) is about 6\% of the input intensity.
    \subsubsection{On-axis efficiencies}

The results of the ESP BL-9 calibration for the ESP on-axis position in horizontal and vertical orientations are shown in Figure 11.
  \begin{figure}[ht]
   \begin{center}
   \begin{tabular}{c}
   \includegraphics[height=6.5 cm]{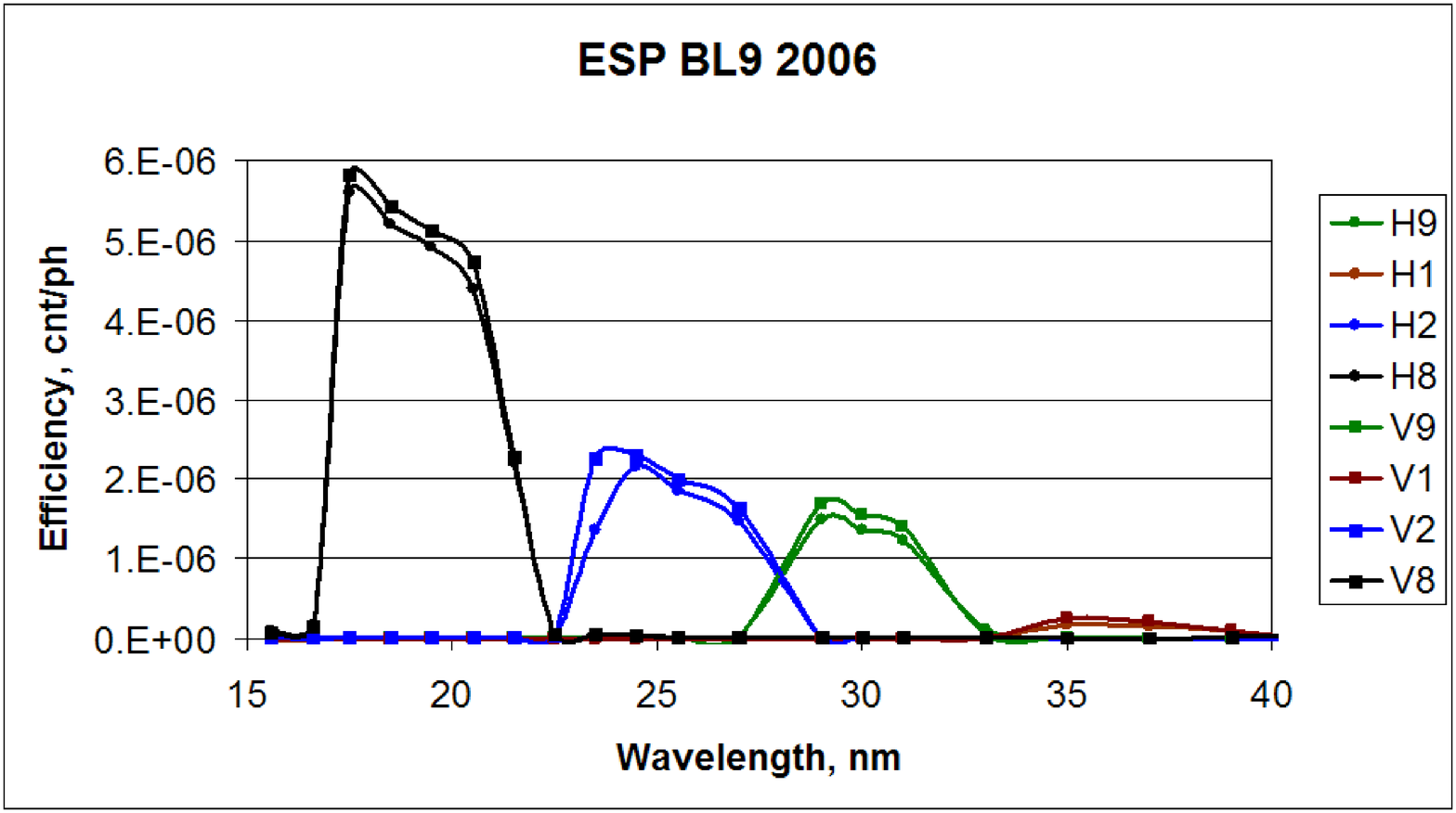}     
   \end{tabular}
   \end{center}
   \caption[Figure 11]
   { \label{fig:Figure 11}
Spectral profiles of ESP first order band efficiencies. From the shorter wavelengths to the longer wavelengths ESP first order efficiencies are shown for Ch8 (19~nm, black), Ch2 (25~nm, blue), Ch9 (30~nm, green), Ch1 (36~nm, brown). Top curves (squares) show profiles for vertical (V) orientation, bottom curves (dots) are for horizontal (H) orientation. The larger efficiency for V than for H is due to the beam polarization}
   \end{figure}
QD efficiency spectral profile is shown in Figure 12. The profile is obtained as a combined result of BL-9 measurements (Figure 9), calculations, and BL-2 calibration.
  \begin{figure}[ht]
   \begin{center}
   \begin{tabular}{c}
   \includegraphics[height=6.5 cm]{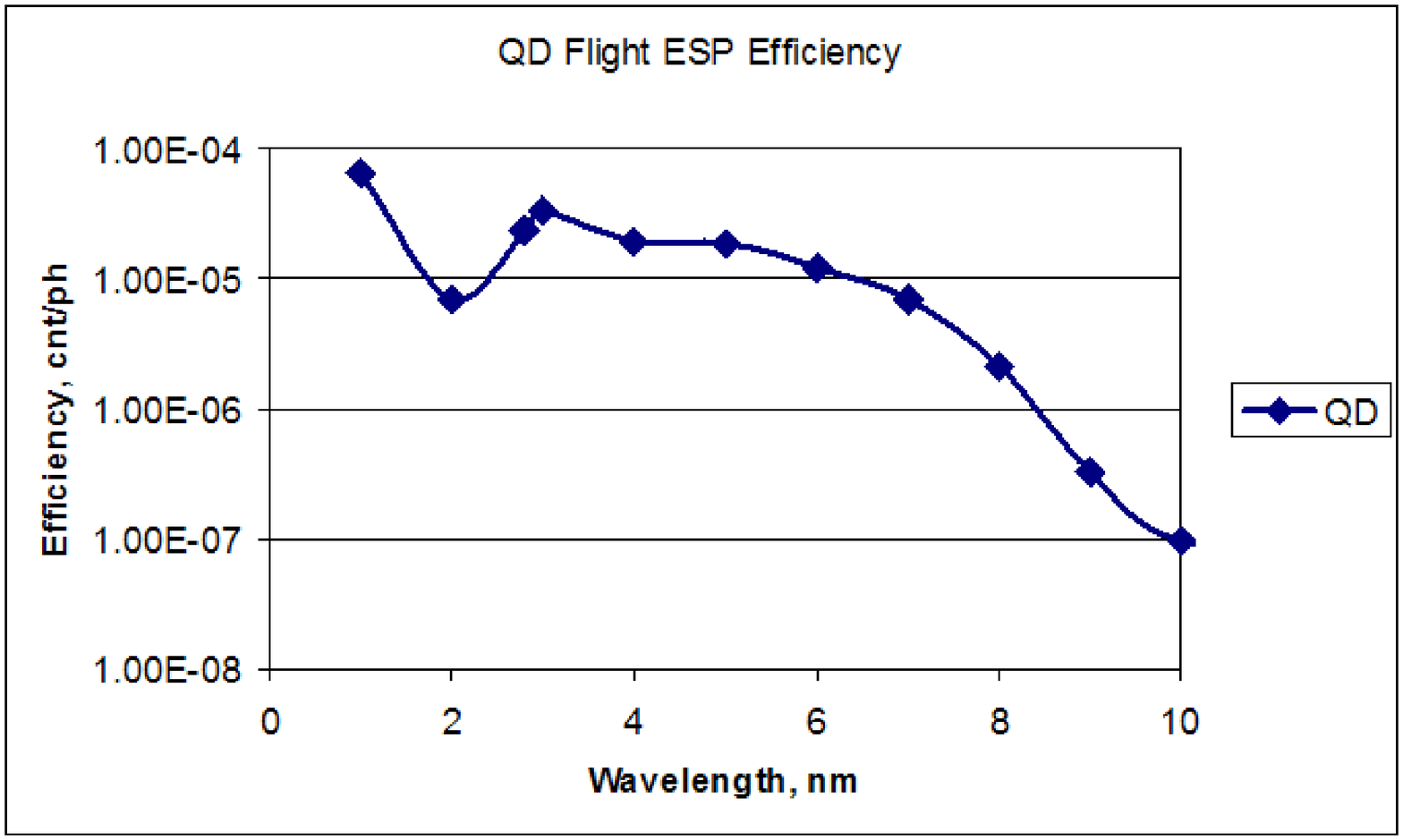}     
   \end{tabular}
   \end{center}
   \caption[Figure 12]
   { \label{fig:Figure 12}
Spectral profile of ESP QD band efficiency. The efficiency drops about three orders of magnitude between 7 nm and 10 nm. This fast decrease of the efficiency below 7 nm has a small effect on calculated zeroth order irradiance as the QD passband does not extend below 7~nm.}
   \end{figure}
 Efficiencies and spectral bandpasses for ESP bands are summarized in Table~3.
\begin{table}[h]
\caption{Efficiencies and bandpasses of the ESP bands }
\begin{tabular}{|c|c|c|c|c|} 
\hline
  ESP  & Maximal (averaged & FWHM  & FWHM, nm & Central wave-\\
  band & for H and V)  & edges, nm & & length, nm   \\
     & efficiency, cnt/ph & & & \\
\hline
Ch1  &  $2.09 \times 10^{-7}$ & 34.0 -- 38.7 & 4.7 & 36.35 \\
Ch2  &  $2.23 \times 10^{-6}$ & 23.1 -- 27.6 & 4.5 & 25.35 \\
Ch8  &  $5.71 \times 10^{-6}$ & 17.2 -- 20.8 & 3.6 & 19.00 \\
Ch9  &  $1.62 \times 10^{-6}$ & 28.0 -- 31.8 & 3.8 & 29.90  \\
QD   &  $6.55 \times 10^{-5}$ & 0.1 -- 7.0 & 6.9 & 3.55 \\
\hline
\end{tabular}
\end{table}

     \subsection{Results from the BL-2 radiometric pre-flight calibration}
     \label{S-BL-2-calibration}

Pre-flight radiometric calibration of ESP-flight (ESPF) was performed on 30 August 2007 at BL-2. ESPF was assembled to EVE and rotated $45^{\circ}$. The goal of that pre-flight radiometric calibration was to determine all characteristics marked as SURF in the second column of the Table 2. The calibration consists of three major parts: alignment to the SURF beam optical axis; tests with ESP tilted positions; tests for ESP optical axis center point.

The alignment starts with vertical and horizontal scans to determine both the intensity center of the beam and the geometrical center of the beam. In the intensity center of the beam the ESP is aligned in both yaw and pitch angles (Figure 7) to have differential signals from the zeroth order QD (bands 4 -- 7) equal to zero. If the differential signal along the dispersion direction of the grating is marked as $X_{d}$ and in the opposite direction as $Y_{d}$, then the equations for these differential signals are:
\begin{equation}    
X_{d} = \frac{Ch6 +Ch7 - Ch4 -Ch5}{Ch4 + Ch5 + Ch6 + Ch7}
\end{equation}
\begin{equation}    
Y_{d} = \frac{Ch5 +Ch6 - Ch4 -Ch7}{Ch4 + Ch5 + Ch6 + Ch7}
\end{equation}
The tests with ESP tilted positions include a number of FOV maps where ESP responses to the tilts are measured as a matrix of tilts in both $\alpha$ and $\beta$ directions (in yaw and pitch), and using cruciform scans in these directions.

\subsubsection{ESP responses for the tilted positions}

The goals of the ESP off-axis calibration were to determine efficiency changes as a function of angle and QD band responses to the tilts.
Some possible misalignment on orbit compared to the on-axis calibration at SURF may occur either as a result of mechanical and/or thermal deformations, or be related to an initial non-perfect co-alignment to the SDO imaging instruments, {\it e.g.}, to AIA. The total misalignment is limited to within $\pm$ 3~arc min. However, ESP may work in a much larger FOV of $\pm$ 2$^{\circ}$ and BL-2 pre-flight calibration included such $\pm$ 2$^{\circ}$ tilts as bi-directional cruciforms. The results of calculated BL-2 irradiance (Equation 1) for $\beta$ scans (along the dispersion) and $\alpha$ scans (in the perpendicular to dispersion direction) are shown in Figure 13, a and b. The plus and minus offsets along $\beta$ axis cause wavelength shifts in different directions for plus and minus first order ESP bands. These shifts increase or decrease calculated irradiance in the vicinity of the optical axis (Figure 13, a). In contrast to these offsets along $\beta$ axis, the offsets in the perpendicular direction, along $\alpha$ axis do not cause wavelength shifts and the changes of calculated irradiance are small for small offsets (Figure 13, b). Larger offsets (more than $\pm$ 0.8$^{\circ}$) cause decrease of irradiance due to vignetting of the beam by detector masks.
     \begin{figure}    
   \centerline{\hspace*{0.015\textwidth}
               \includegraphics[width=0.4\textwidth,clip=]{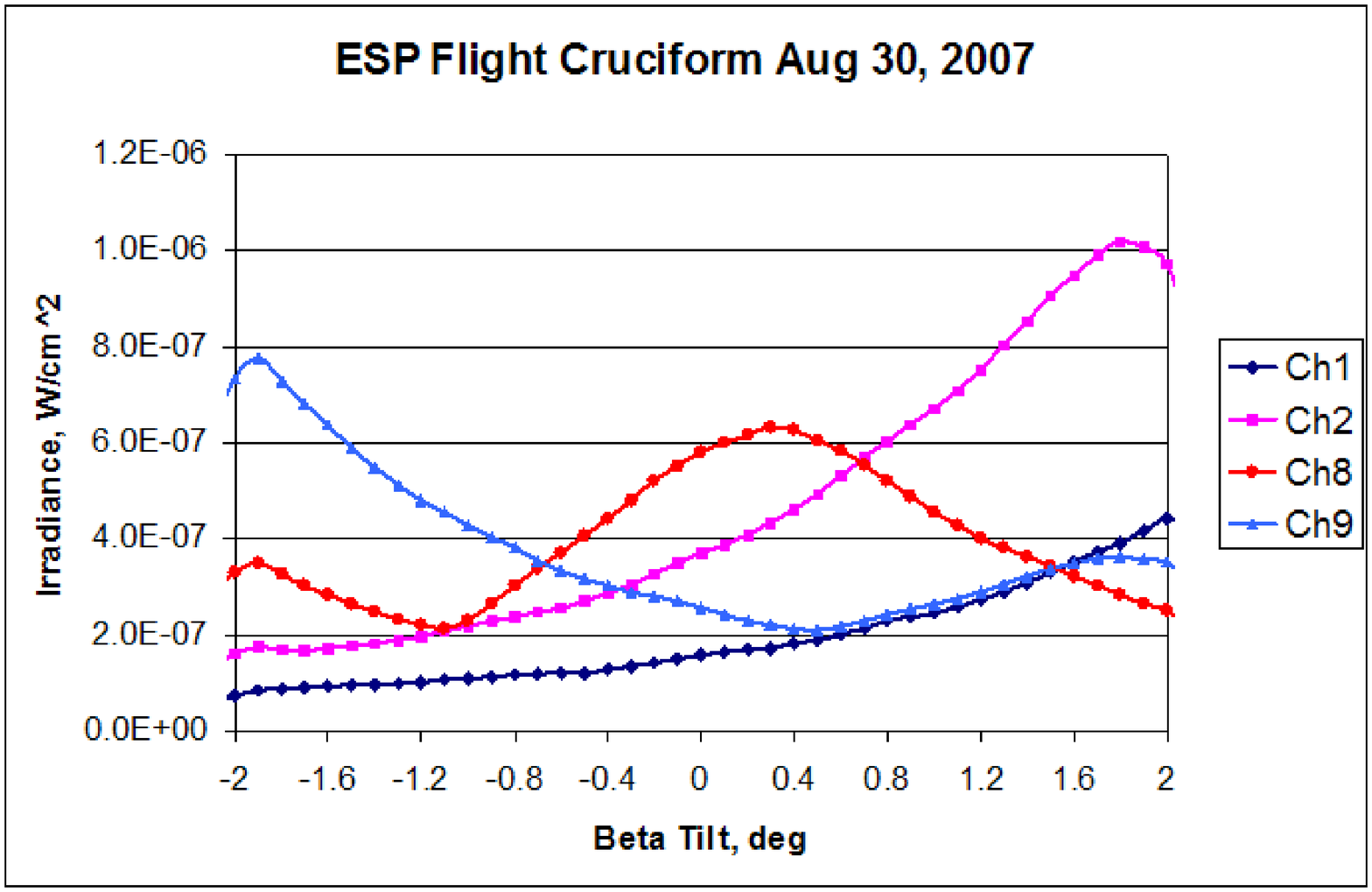}
               \hspace*{-0.01\textwidth}
               \includegraphics[width=0.4\textwidth,clip=]{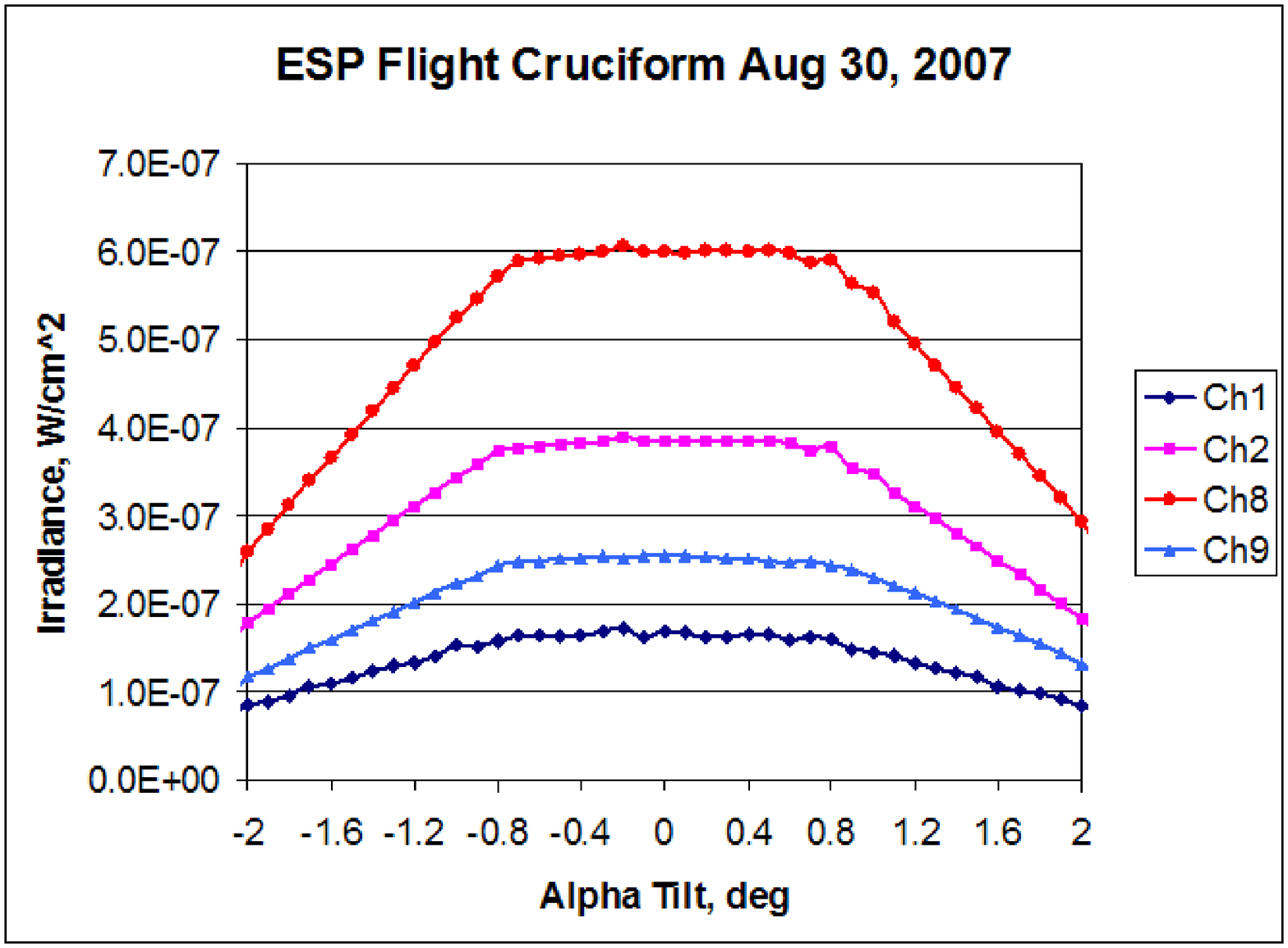}
              }
     \vspace{-0.31\textwidth}   
     \centerline{\Large \bf     
      \hspace{0.1 \textwidth}  \color{black}{(a)}
      \hspace{0.3\textwidth}  \color{black}{(b)}
         \hfill}
     \vspace{0.27\textwidth}    
   \centerline{\hspace*{0.015\textwidth}
               \includegraphics[width=0.4\textwidth,clip=]{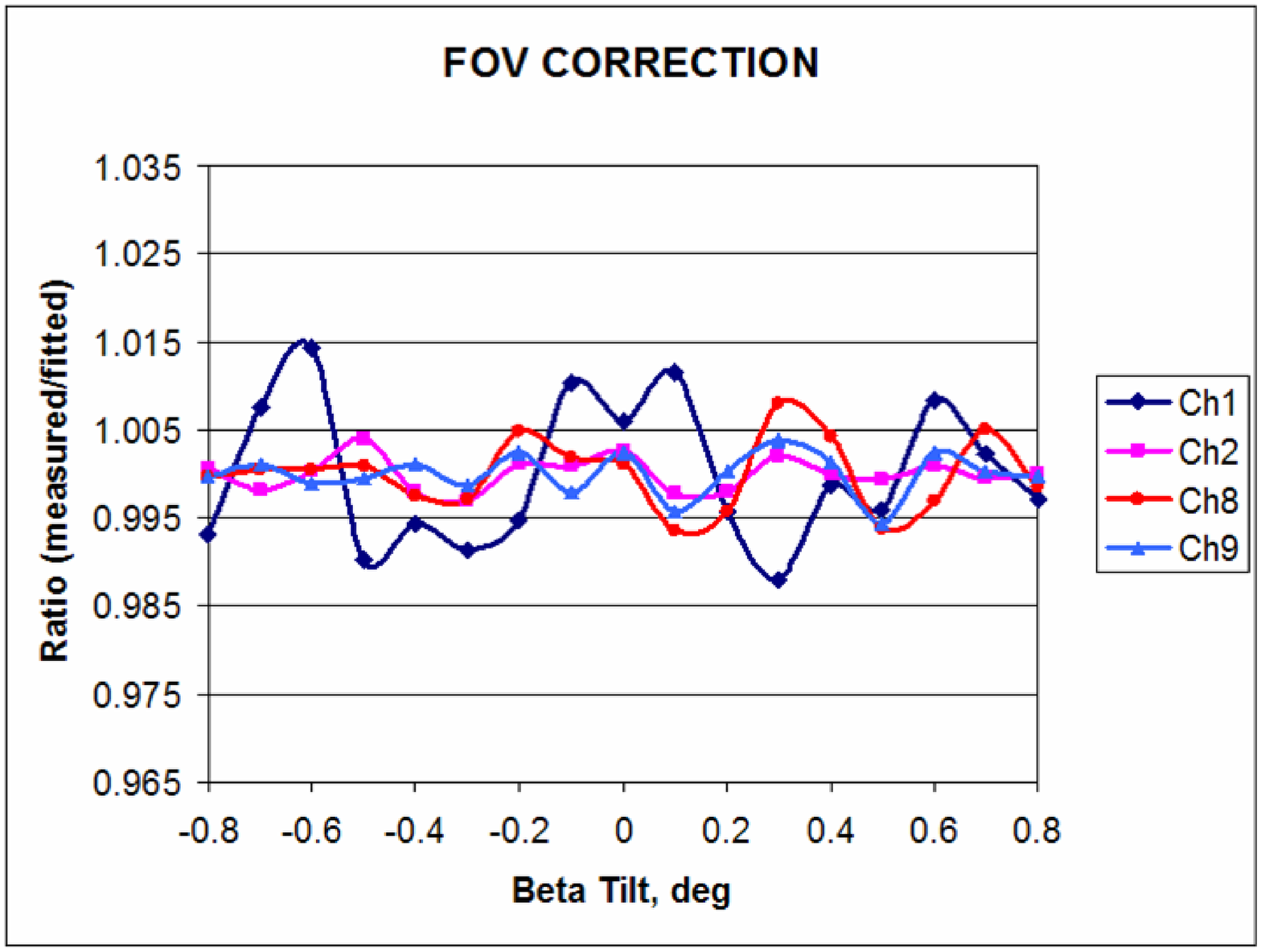}
               \hspace*{-0.01\textwidth}
               \includegraphics[width=0.4\textwidth,clip=]{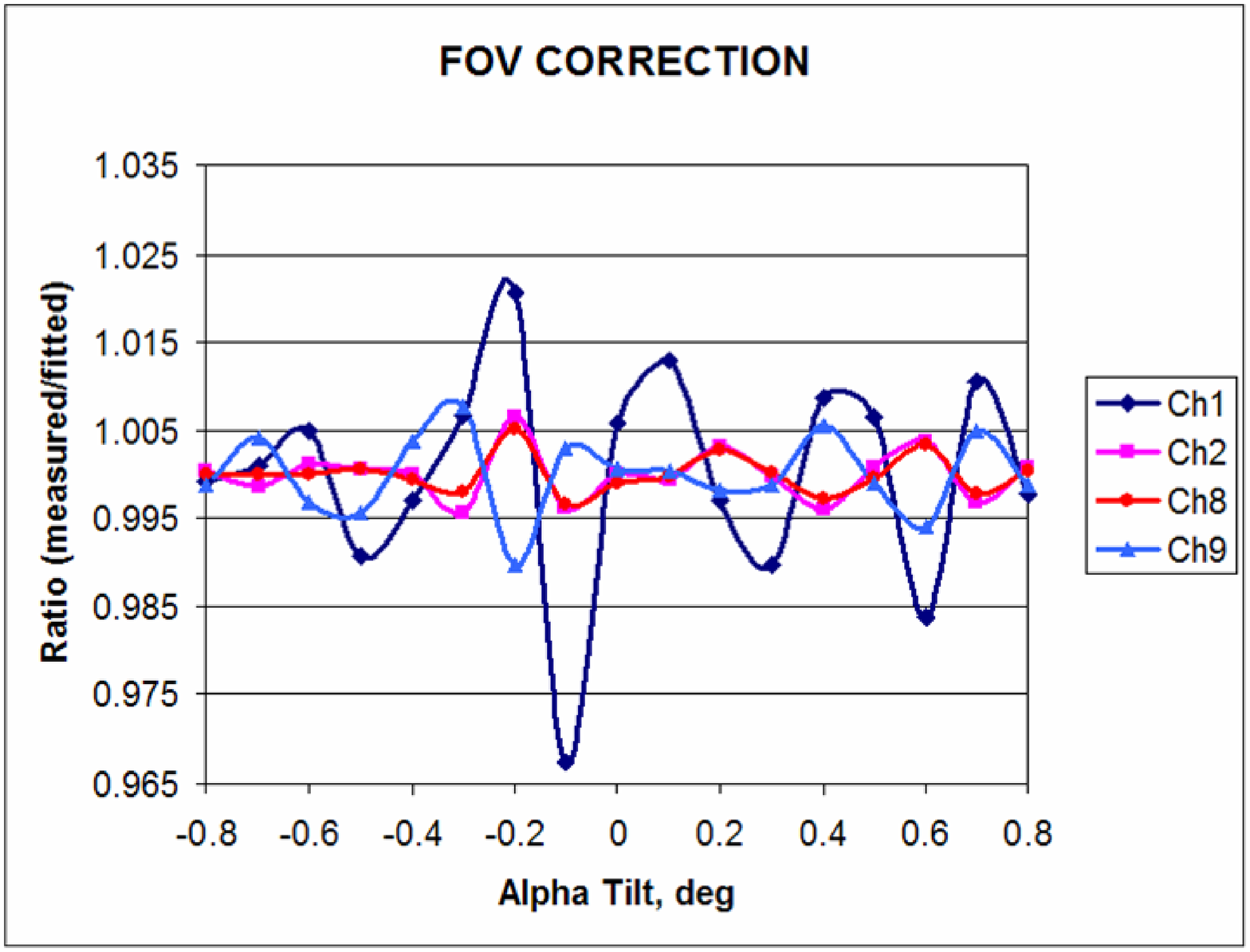}
              }
     \vspace{-0.31\textwidth}   
     \centerline{\Large \bf     
      \hspace{0.1 \textwidth} \color{black}{(c)}
      \hspace{0.3\textwidth}  \color{black}{(d)}
         \hfill}
     \vspace{0.28\textwidth}    
\caption{Calculated irradiance (Equation 1) for the full FOV of ($\pm$ 2$^{\circ}$) from the BL-2 cruciform in the dispersion direction (a). Same for the perpendicular to the dispersion direction (b). A result of applying correction curves to the cruciform profiles is shown on the bottom panels. The level of the correction is shown as a ratio between the measured values for tilted positions and the fitted values (from the curves) for the dispersion direction (c), and for the perpendicular direction, (d). The correction fit was applied to the FOV of ($\pm$ 0.8$^{\circ}$), 1.5 times larger than the solar angular size. }
   \label{F-cruciform}
   \end{figure}

The Figure 13 bottom panels, c and d show ratios between measured and fitted curves as non perfectly compensated fluctuations with amplitudes of up to $\pm$ 1.5\% (c: dispersion direction), and $\pm$ 2.7 \% (d: perpendicular to the dispersion direction).

Another reason to calibrate ESP in tilted positions is to determine the sensitivity of the zeroth order bands to the tilts. The information about this sensitivity is important for establishing tilts based on the differential signals (Equations 16, 17). This zeroth order sensitivity will be used on orbit to determine the amount of misalignment and to calculate the position of a solar flare on the disk with near real time information. The QD sensitivity to the tilts is shown in Figure 14.
 \begin{figure}    
   \centerline{\hspace*{0.015\textwidth}
               \includegraphics[width=0.51\textwidth,clip=]{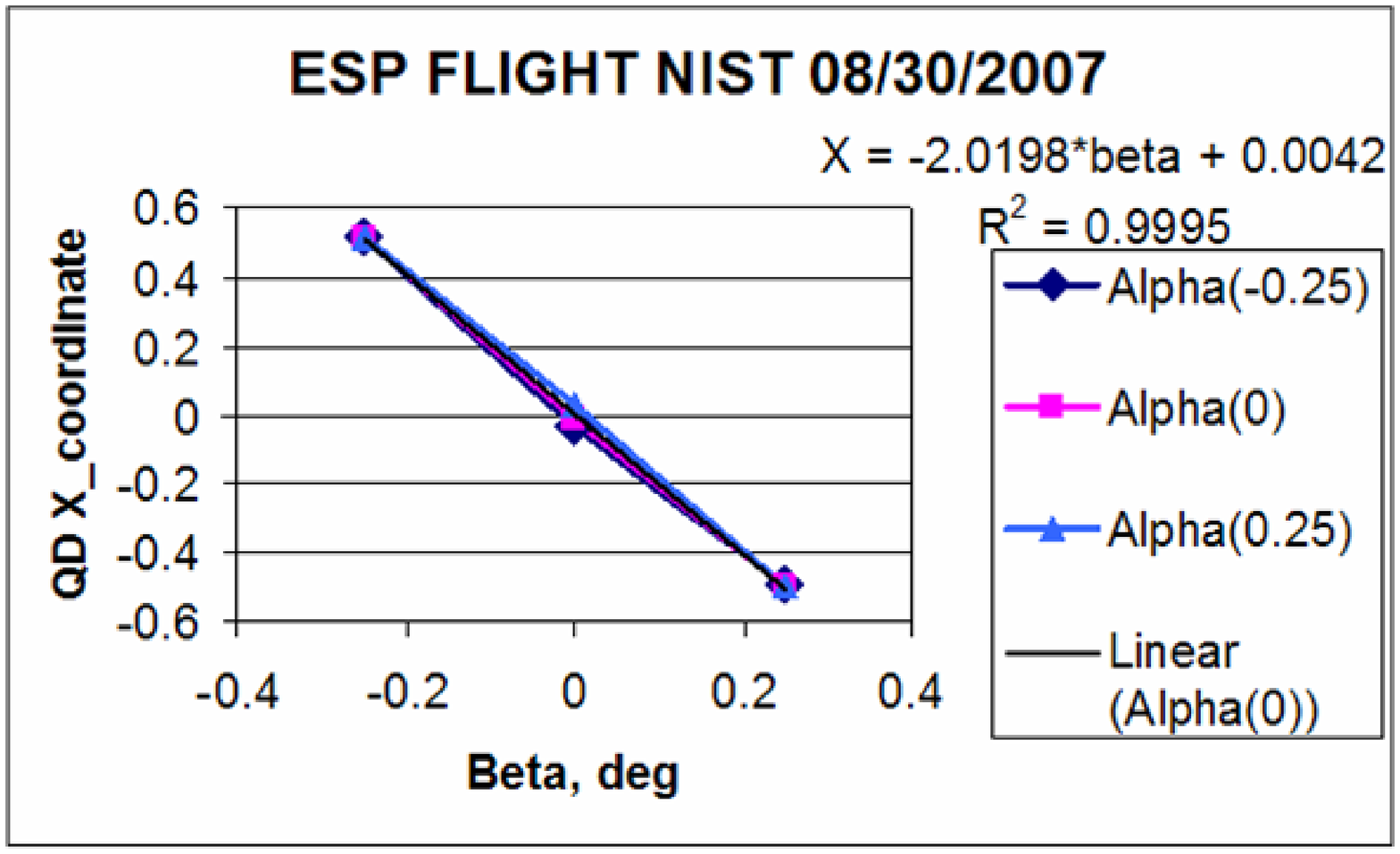}
               \hspace*{-0.03\textwidth}
               \includegraphics[width=0.48\textwidth,clip=]{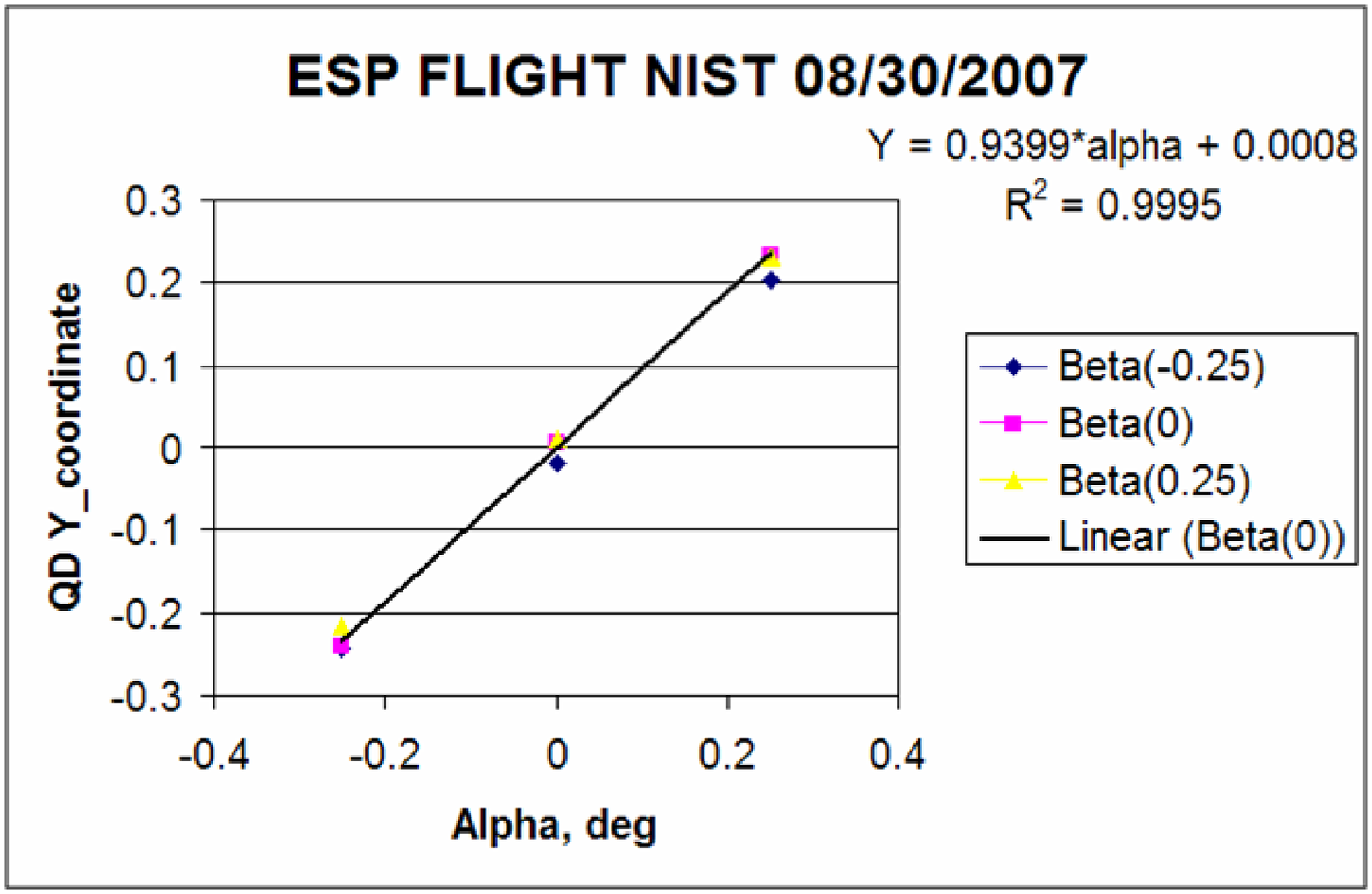}
              }
     \vspace{-0.3\textwidth}   
     \centerline{\Large \bf     
      \hspace{0.02 \textwidth}  \color{black}{(a)}
      \hspace{0.4\textwidth}  \color{black}{(b)}
         \hfill}
     \vspace{0.25\textwidth}    
     \caption{(a) The sensitivity of QD bands to tilts along the dispersion direction. (b) Same for the perpendicular to the dispersion direction. The sensitivity in each direction is a linear function within the tilts of $\pm$ 0.25$^{\circ}$. The sensitivity in the vertical (alpha) direction is about two times lower than the sensitivity in the dispersion (beta) direction.  }
   \label{F-QD-tilts}
   \end{figure}

  Determined sensitivities (Equations 19, 20) to the tilts within $\pm$ 0.25$^{\circ}$ are:
  \begin{equation}  
X_{d} = -2.02 \times \beta + 0.0042
\end{equation}
\begin{equation}    
Y_{d} = 0.94 \times \alpha + 0.0008
\end{equation}

\subsubsection{BL-2 Calibration for ESP FOV Maps}

Calibration for the ESP in tilted positions includes two-dimensional, nine-point, FOV maps with eight points in which either alpha, delta, or both are tilted to $\pm$ 0.25$^{\circ}$ and one point for the center of the FOV. Such maps show the angular change of $R_{i}(\lambda,\alpha,\beta)$ (1). The results for the plus (Ch1) and minus (Ch9) first diffraction order bands are shown in Figure 15.
\begin{figure}    
   \centerline{\hspace*{0.015\textwidth}
               \includegraphics[width=0.51\textwidth,clip=]{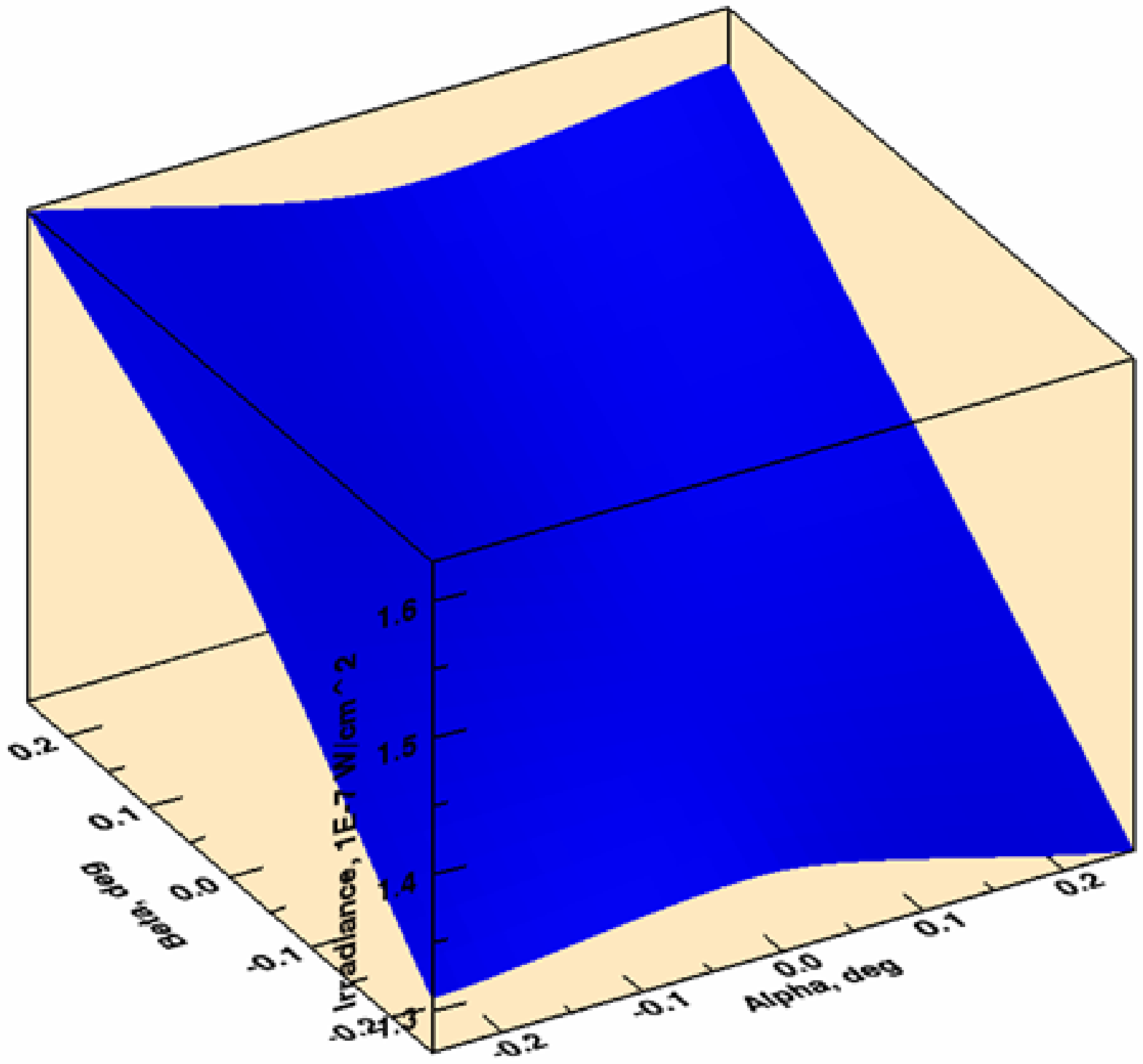}
               \hspace*{-0.03\textwidth}
               \includegraphics[width=0.48\textwidth,clip=]{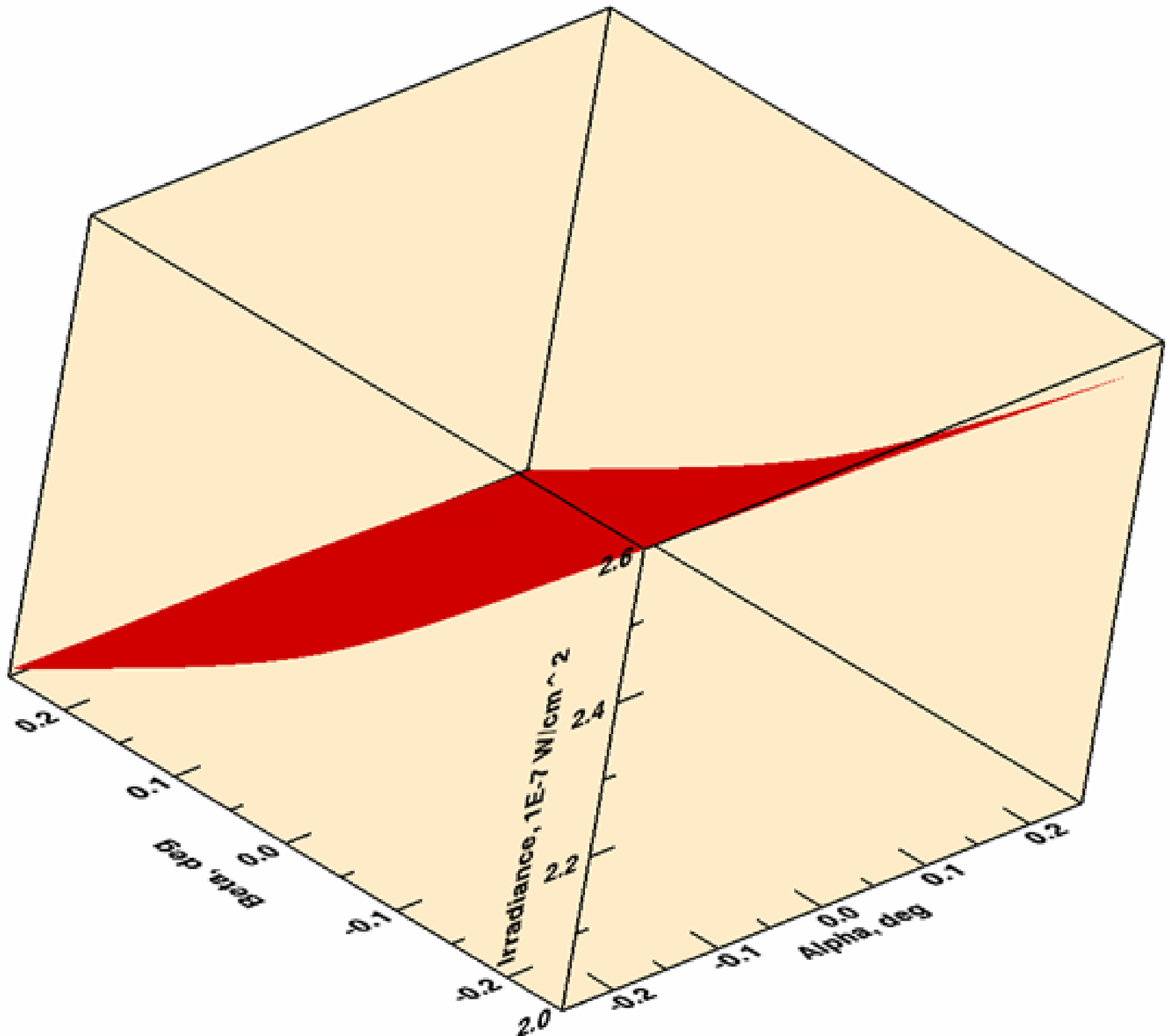}
              }
     \vspace{-0.45\textwidth}   
     \centerline{\Large \bf     
      \hspace{0.02 \textwidth}  \color{black}{(a)}
      \hspace{0.4\textwidth}  \color{black}{(b)}
         \hfill}
     \vspace{0.4\textwidth}    
     \caption{(a) An example of the change of calculated irradiance (in units of 1E-7 W/$cm^{2}$  in the $\pm$ 0.25$^{\circ}$ FOV. Note the significantly larger change of the irradiance in the grating dispersion (Beta) direction than in Alpha direction related to the shift of the diffraction maximum with the change of $\theta$ (15). (b) Same for Ch~9. Note that the irradiance change for plus and minus Beta is opposite to the Ch~1 (a) trend. The changes of the irradiance for both a) and b) in the Alpha direction are small.}
   \label{F-maps}
   \end{figure}
ESP calibration for the FOV maps with two-dimensional tilts provides additional information to the single Alpha or Beta cruciform tests (see above) about changes of the measured irradiance in the ESP $\pm$ 0.25$^{\circ}$ FOV.

\subsubsection{BL-2 Calibration for ESP On-Axis Position}

ESP BL-2 calibration in the on-axis position includes calibration for the set of three aluminum filters in the filter wheel (Figure 1) and calibration with different synchrotron beam energies, 380~MeV (a primary energy for calibration), 331~MeV, 285~MeV, 229~MeV, 183~MeV, and 140~MeV. These different energies shift the spectral distribution to provide a wide variety of emitted irradiation, from X-ray to visible light, suitable to calibrate the scientific bands and to permit a determination of the amount of higher order diffraction contamination (order sorting). The results of ESPF pre-flight calibration (30 August 2007) for the beam energy of 140~MeV, which provides the spectral distribution with very low higher order contamination, is given in Table 4. It shows a comparison of BL-2 irradiance which was entering each ESPF spectral band and the irradiance measured by the ESPF. For the zeroth order QD band which is free of higher order contamination the comparison is given for the beam energy of 380~MeV.
\begin{table}[h]
\caption{A comparison of the BL-2 140~MeV irradiance that entered the ESPF bands (second column) and the measured irradiance (third column). The fourth column shows the spectral bands and the fifth column gives relative error for this comparison. QD comparison was performed with the irradiance calculated for the beam with energy of 380~MeV.}
\begin{tabular}{|c|c|c|c|c|} 
\hline
  ESP  & Input & Measured by ESPF, & Spectral band, & $\Delta, \% $\\
  band & irradiance, W/cm$^{2}$ & irradiance, W/cm$^{2}$ & nm & \\
\hline
Ch1  &  $1.83 \times 10^{-6}$ & $1.83 \times 10^{-6}$& 34.3 -- 38.5 & 0.0 \\
Ch2  &  $8.16 \times 10^{-7}$ & $8.25 \times 10^{-7}$& 23.4 -- 28.1 & 1.1\\
Ch8  &  $1.69 \times 10^{-7}$ & $1.70 \times 10^{-7}$ & 17.5 -- 21.1 & 0.6 \\
Ch9  &  $1.37 \times 10^{-6}$ & $1.38 \times 10^{-6}$ & 28.0 -- 32.7 & 0.7  \\
QD  &  $1.13 \times 10^{-5}$ & $1.12 \times 10^{-5}$ & 1.0 -- 7.0 & 0.9  \\
\hline
\end{tabular}
\end{table}
All three aluminum filters on the ESP filter wheel show very similar transmission with deviations less than $\pm 2.5 \%$. Figure 16 shows the variation in efficiencies among these filters (1 through 3).
  \begin{figure}[ht]
   \begin{center}
   \begin{tabular}{c}
   \includegraphics[height=6.5 cm]{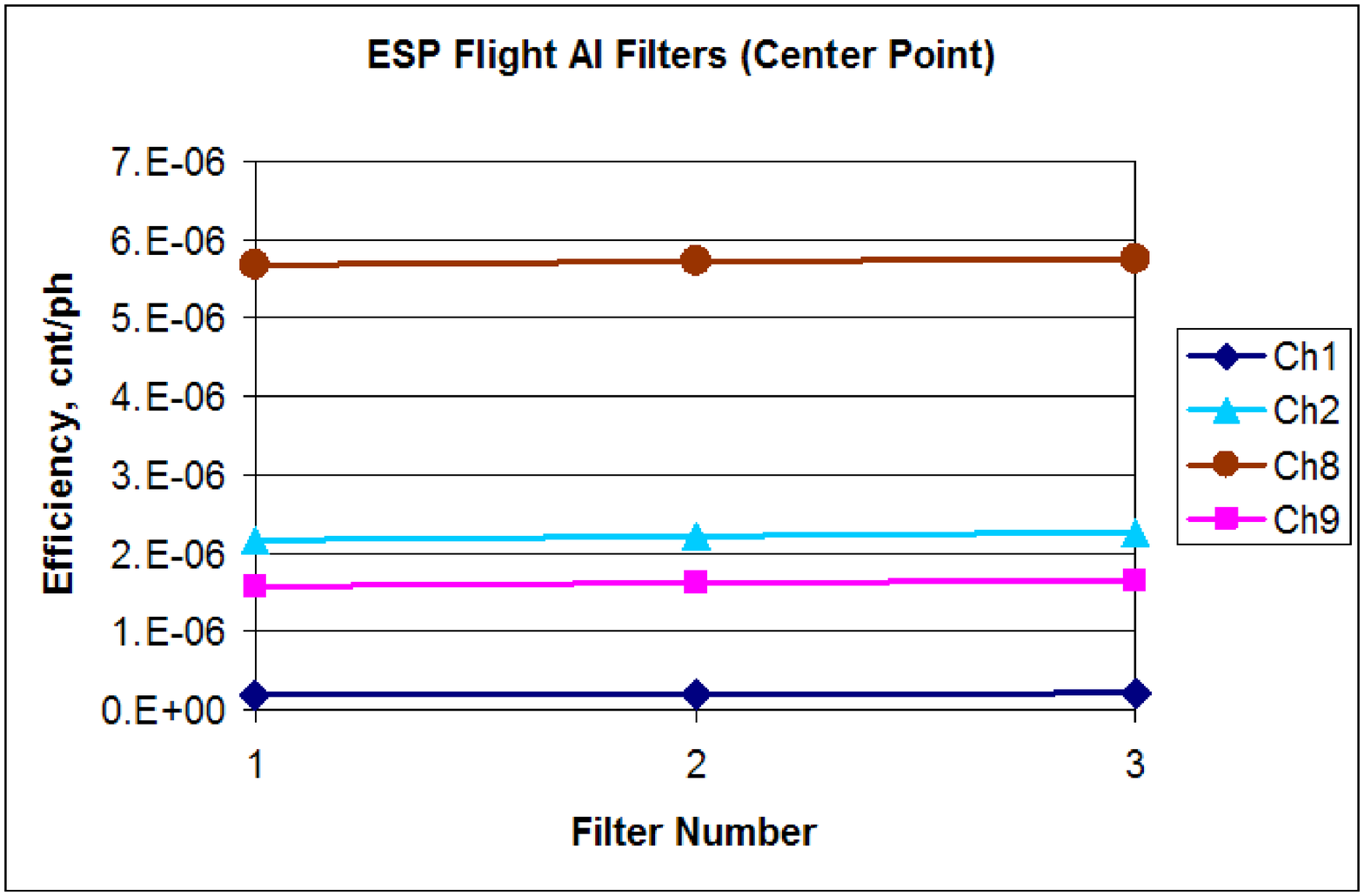}           
   \end{tabular}
   \end{center}
   \caption[Figure 16]
   { \label{fig:Figure 16}
ESP first order band efficiencies for three aluminum filters installed in the filter wheel of the flight ESP. The measured changes of the efficiencies related to a small difference in transmission for these Al filters are small, about $\pm 2.5 \%$.  }
   \end{figure}

\subsubsection{ESP sensitivity to the higher orders (BL-2 order sorting test)}

The BL-2 radiometric calibration includes a special test called `Order Sorting'. This test consists of a number of calibrations at different electron beam energies, and thus varies the spectral content of the photon beam with which the ESP entrance aperture is illuminated. An example of BL-2 spectra for beam energies of 380 MeV, 331 MeV, and 140 MeV is shown in Figure 17.
  \begin{figure}[ht]
   \begin{center}
   \begin{tabular}{c}
   \includegraphics[height=6.5 cm]{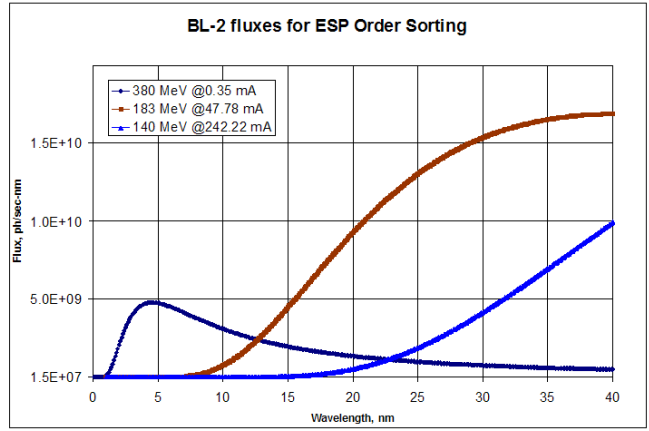}           
   \end{tabular}
   \end{center}
   \caption[Figure 17]
   { \label{fig:Figure 17}
An example of BL-2 fluxes for the beam energies of 380, 183, and 140 MeV, at beam currents of 0.35, 47.8, and 242.2 mA, correspondingly. }
   \end{figure}

Figure 17 shows that a calibration with the beam energy of 140 MeV allows for the detection and measurement of the input irradiance mainly by the first order bandpasses. For example, the flux for the second order of Ch9 at about 15~nm is more than two orders of magnitude lower than that of the first order. These low energy beams (140~MeV and 183~MeV) allow us to calibrate ESP with a minimal amount of higher order irradiation. In contrast, the QD calibration use beams with high energy, {\it e.g.} 380 MeV and 331 MeV. These higher energy beams provide sufficient signal to the QD zeroth order channels (0.1 to 7.0 nm) and allow precise alignment of the ESP to the beam, which is important during the calibration tests.
The results of the Order Sorting test are shown in Figure 18.
  \begin{figure}[ht]
   \begin{center}
   \begin{tabular}{c}
   \includegraphics[height=6.5 cm]{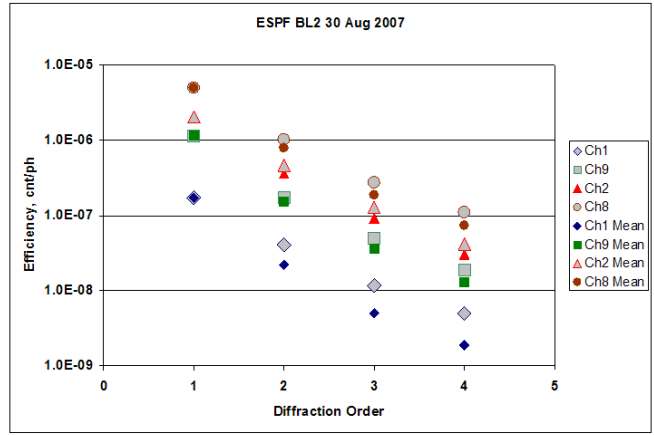}           
   \end{tabular}
   \end{center}
   \caption[Figure 18]
   { \label{fig:Figure 18}
ESP efficiencies for the first order channels (left side points) and range of the efficiency changes, between gray and colored points for the second, third, and fourth orders. The points were determined using two different modeling approaches.}
   \end{figure}

Figure 18 shows two sets of points (grey and filled with colors for Ch1 (blue diamonds), Ch2 (red triangles), Ch8 (brown circles), and Ch9 (green squares) calculated for the first through the fourth order using beam energies of 140 MeV, 183 MeV, and 380 MeV. The efficiencies marked with the grey points were modeled assuming no change of the mean efficiency $Eff_{i}$ for each order $i$ with different beam energies $E$ and beam currents $I_{E}$. The mean efficiency was determined as $(1/n_{\lambda}\sum_{\lambda1}^{\lambda2}Eff(\lambda)$. Thus, effective counts $C_{eff}$ are the mean efficiency multiplied by the integrated flux $ \sum_{\lambda1i}^{\lambda2i}\Phi$:
 \begin{equation}   
C_{eff} = \sum_{i=1}^{i=4}Eff_{i}(\sum_{\lambda1i}^{\lambda2i}\Phi_{E}(\lambda) \times I_{E} \times \Delta\lambda),
\end{equation}
where $\lambda1i$, $\lambda2i$ are the edges of the bandpass for the order $i$.

The efficiencies marked with color filled points are based on another approach which allows a change of the modeled efficiency due to flux shape change inside the channel's bandpass. The change of the spectral flux profile related to the switch from one energy to another (see Figure 17) shifts the mean efficiency for the channel, order, and the beam energy, allowing more accurate modeling of the effective counts measured by the channel. Each higher order efficiency color point in Figure 18 represents the averaged efficiency for the three beam energies used. The channel's efficiency (one averaged number for all orders) determined as measured effective counts $C_{eff}$ divided by the flux $\Phi(E,i)$ integrated over all order bandpasses $\sum_{i,\lambda}\Phi$ for this approach is compared to the sum of the mean efficiencies for each beam energy $E$ as:
 \begin{equation}   
\frac{C_{eff}(Ch,E)}{\Phi(E,i)} = \sum_{i=1}^{i=4}Eff(Ch,E,i) \times \frac{\Phi(Ch,E,i)}{\Phi(E,i)},
\end{equation}
The right side of Equation 22 is the sum of efficiencies weighted by the ratio of the fluxes and, thus, sensitive to the shape of the flux profile in each higher order bandpass.

The higher order counts for the BL-2 ESP calibration is shown in Figure 19. The ratio between the measured effective counts and the counts extracted in the first order bandpass is illustrated.
  \begin{figure}[ht]
   \begin{center}
   \begin{tabular}{c}
   \includegraphics[height=6.5 cm]{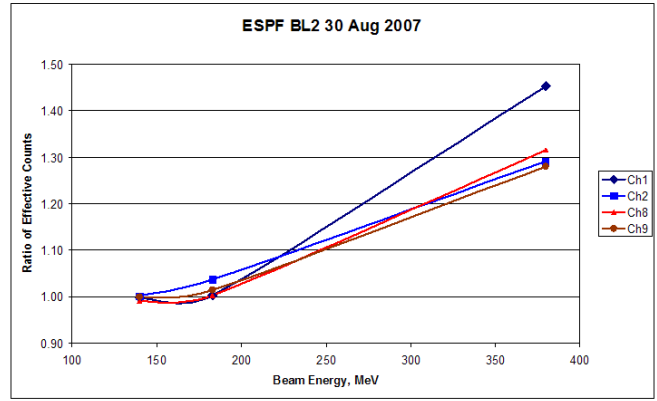}           
   \end{tabular}
   \end{center}
   \caption[Figure 19]
   { \label{fig:Figure 19}
Amount of higher order irradiance for three BL-2 beam energies, 140, 183, and 380~MeV is shown for the ESP first order channels as a ratio between the measured effective counts and the counts extracted from the first order bandpass. }
   \end{figure}

The ratio of 1.0 is within one percent for the beam energy of 140~Mev. It starts to be higher for the beam energy of 183~Mev and reaches the maximal ratio of 1.45 for Ch1 at the beam energy of 380~MeV. These ratios will be significantly lower for solar measurements with narrow spectral lines in contrast to the more continuous profiles characteristic of synchrotron radiation (Figure 17). Analyses for the higher order contribution during solar measurements with SOHO/SEM showed that for the He II (30.4~nm) solar irradiance for quiet and intermediate activity these contributions are smaller than 10\%.

 \section{A Comparison of ESP Measured Irradiance from the Sounding Rocket Flight with other Measurements }
     \label{S-Irradiance-Comparison}

  A sounding rocket flight (NASA Rocket 36.240) with the EVE suite of channels was flown on 14 April 2008. The ESP rocket instrument (ESPR) collected data in both first order and zeroth order bands. ESP operational software was used to convert the count rates measured near apogee into solar irradiance according to Equation (1). The calculated irradiance was corrected for the Earth's atmosphere absorption (which is wavelength, date, and altitude dependent) using the MSIS \cite{Hedin87, Picone03} atmosphere model. Figure 20 shows an example of the original count rates before subtracting dark currents for two ESPR bands, Ch8 and Ch9.
  \begin{figure}[ht]
   \begin{center}
   \begin{tabular}{c}
   \includegraphics[height=6.5 cm]{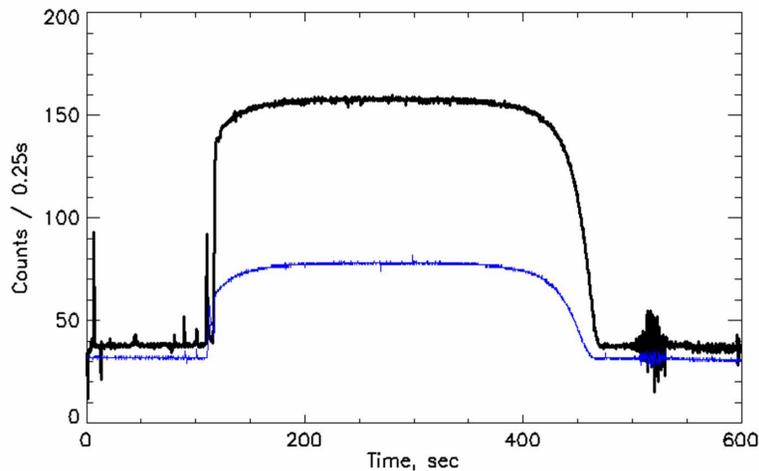}          
   \end{tabular}
   \end{center}
   \caption[Figure 20]
   { \label{fig:Figure 20}
An example of observing profiles (detector counts {\it vs.} time) for two ESPR bands (Ch8 and Ch9) obtained during EVE sounding rocket flight 36.240 flown on April 14, 2008. Black line shows Ch8 (19~nm) observing profile, blue line is for Ch9 (30.4~nm) data. The count rate on the plot includes dark counts (about 37.8 cnt/0.25~sec for Ch8 and 31.9 cnt/0.25~sec for Ch9). The time starts as the rocket was launched. The apogee point with altitude of about 285~km was achieved 275~sec into the flight. }
   \end{figure}
Solar irradiance determined from the sounding rocket flight ESPR measurements in the first and zeroth order bands was compared to the irradiance obtained from the EVE/MEGS spectra integrated over the same wavelengths. The results of this comparison are shown in Table 5.
\begin{table}[h]
\caption{A comparison of ESPR and MEGS absolute solar irradiance determined from the 14 April 2008 sounding rocket flight. MEGS spectral irradiance was integrated over the ESPR wavelengths (fourth column) determined using the convolution as shown in Equation 11 (see also Figure 2 for Ch8 spectral band). The last column shows relative difference between ESP and MEGS irradiance.}
\begin{tabular}{|c|c|c|c|c|} 
\hline
  ESP  & ESP irradiance, & MEGS irradiance, & Spectral band, nm & $\Delta, \% $\\
  band & W/m$^{2}$ & W/m$^{2}$ & \\
\hline
Ch1  &  $1.28 \times 10^{-4}$ & $1.32 \times 10^{-4}$& 33.0 -- 38.55 & 3.0 \\
Ch2  &  N/A & N/A & N/A & N/A\\
Ch8  &  $4.65 \times 10^{-4}$ & $4.65 \times 10^{-4}$ & 14.5 -- 22.2 & 0.0 \\
Ch9  &  $5.20 \times 10^{-4}$ & $5.28 \times 10^{-4}$ & 26.7 -- 33.8 & 1.5  \\
QD  &  $9.86 \times 10^{-5}$ & $9.77 \times 10^{-5}$ & 0.1 -- 7.0 & 0.9  \\
\hline
\end{tabular}
\end{table}
At the time of the sounding rocket flight ESPR Ch2 was not working due to low shunt resistance of the diode detector. This detector was replaced in December 2008, and currently the ESPR instrument is fully operational and its bands are calibrated. The results of the BL-2 (23 January 2009) radiometric calibration for the updated ESPR are shown in Table 6. Table 6 compares ESPR input and measured irradiance similar to the comparison given in Table 4 for the ESPF.
\begin{table}[h]
\caption{A comparison of BL-2 irradiance that entered the ESPR bands (second column) for the beam energy of 140~MeV with the irradiance measured by the ESPR (third column). The fourth column shows the spectral bands and the fifth column gives relative error of this comparison. QD comparison was performed with the irradiance calculated for the beam with energy of 380~MeV. }
\begin{tabular}{|c|c|c|c|c|} 
\hline
  ESP  & Input & Measured by ESPR & Spectral band, & $\Delta, \% $\\
  band & irradiance, W/cm$^{2}$ & irradiance, W/cm$^{2}$ & nm & \\
\hline
Ch1  &  $2.22 \times 10^{-6}$ & $2.24 \times 10^{-6}$& 34.2 -- 38.5 & 0.9 \\
Ch2  &  $8.02 \times 10^{-7}$ & $8.12 \times 10^{-7}$& 23.3 -- 27.4 & 1.2\\
Ch8  &  $1.79 \times 10^{-7}$ & $1.85 \times 10^{-7}$ & 17.5 -- 20.9 & 3.4 \\
Ch9  &  $1.63 \times 10^{-6}$ & $1.65 \times 10^{-6}$ & 28.0 -- 32.7 & 1.2  \\
QD  &  $1.01 \times 10^{-5}$ & $1.01 \times 10^{-5}$ & 1.0 -- 7.0 & 0.0  \\
\hline
\end{tabular}
\end{table}
ESP Ch9 (30.4~nm) flux for the 14 April 2008 measurements was also compared to the SOHO/SEM first order flux (26 to 34~nm). To make this comparison, ESP Ch9 flux was re-calculated for the SEM first order bandpass and plotted in Figure 21 as a brown circle. Figure 21 shows SEM calibrated absolute first order flux in the units of ph/cm$^{2}$ sec (blue points) with over-plotted sounding rocket data from both the SEM clone instrument (red squares) and the Rare Gas Ionization Cell (RGIC; black triangles).
  \begin{figure}[ht]
   \begin{center}
   \begin{tabular}{c}
   \includegraphics[height=8.5 cm]{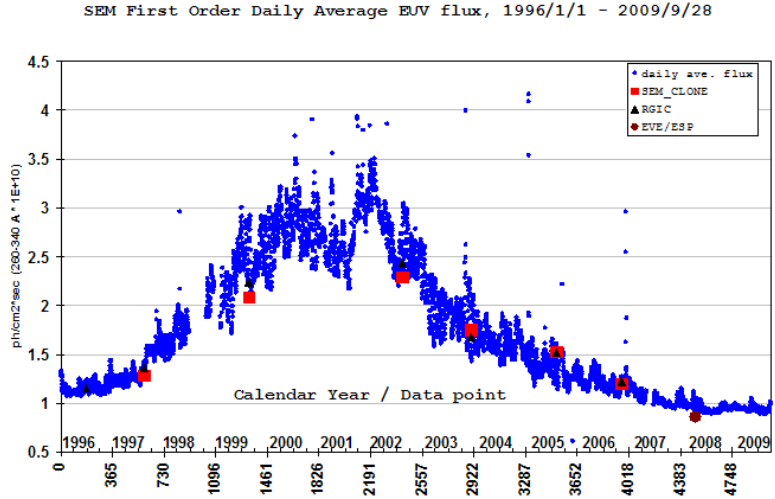}
   \end{tabular}
   \end{center}
   \caption[Figure 21]
   { \label{fig:Figure 21}
A comparison of SOHO/SEM first order absolute EUV flux data for 26 to 34 nm (blue diamonds) with the SEM clone sounding rocket data in the same wavelength bandpass (red squares), RGIC data converted to the SEM bandpass (black triangles) from the same sounding rocket under-flights, and from EVE/ESP (Ch9 at 28.0 to 31.8 nm) sounding rocket flight (brown circle). Solar spectral variability in the SOHO/SEM bandpass was calculated based on the SOLERS-22 solar model data \cite {Woods98b} used for SOHO/SEM calibration since 1996.}
   \end{figure}
SOHO/SEM data show that solar minimum occurred by the end of 2008.
ESP sounding rocket flight data are in good agreement with both EVE/MEGS and SOHO/SEM data.

 \section{Concluding Remarks }
     \label{S-Conclusions}
As part of the SDO/EVE suite of channels, ESP will provide highly stable and accurate measurements of absolute solar irradiance in five EUV wavelength bands. Its high temporal cadence, low latency, and spectral overlap with the other SDO instruments, will provide important details of the dynamics of rapidly changing irradiance related to impulsive phases of solar flares. An algorithm to convert measured count rates and other ESP data into solar irradiance was described. ESP has many significant design improvements over its SOHO/SEM predecessor, including the possibility of measuring on orbit changes of dark currents, electronics gain changes, contamination degradation and pin-holes of thin-film filters, visible light scatter, and energetic particles background. All optical components of ESP (diodes, filters, grating) were also separately tested and calibrated prior to an end-to-end calibration. The ESP was calibrated at SURF BL-9 using quasi-monochromatic wavelengths, and at BL-2 to obtain a radiometric calibration. Efficiency profiles determined during BL-9 calibration were used as reference data for calculation of irradiance in each band during BL-2 calibration. Measurements of solar irradiance from ESPR during the EVE sounding rocket flight of 14 April 2008 were compared with corresponding EVE/MEGS spectra and SOHO/SEM irradiance measurements. These comparisons show good agreement. Both the ESP flight and rocket instruments are fully calibrated and ready for solar measurements.
\begin{acks}
 The authors thank Frank Eparvier, Mike Anfinson, Rick Kohnert, Greg Ucker, Don Woodraska, Karen Bryant, Gail Tate, Matt Triplett, and the rest of the LASP EVE Team at the University of Colorado at Boulder for their many contributions during ground tests of the ESP. We also thank Don McMullin of the Space Systems Research Corporation for his discussions and long-term support of this mission, and the  NIST SURF Team: Rob Vest, Mitch Furst, Alex Farrell, and Charles Tarrio for support of ESP calibration at BL-9 and BL-2 and diffraction grating transmission measurements at the SURF reflectometer facility. This work was supported by the University of Colorado award 153-5979.
\end{acks}


\end{article}

\end{document}